\documentclass[a4paper,12pt]{article}
\usepackage[english]{babel}
\usepackage{epsf,epsfig,graphicx}
\usepackage{amssymb}
\usepackage{amsmath}
\usepackage{amsthm}
\usepackage{multicol}
\usepackage[hang]{subfigure}
\usepackage{cite}
\usepackage{geometry}
\makeatletter
\newcommand{\hm}[1]{#1
\nobreak\discretionary{}{\hbox{\ensuremath{#1}}}{}} \makeatother
\begin{document}
\begin{center}

\textbf{Nonperturbative region of effective strong coupling}

\vskip 5mm

{Viktor Andreev}\\

\vskip 5mm Gomel State University, Gomel, Belarus
\end{center}

\begin{abstract}

\vskip 3mm

\noindent In the framework of Poincar\'{e} covariant quark model
the behavior of running coupling constant
$\alpha_{s}\left(Q^{2}\right)$   is considered in $Q < 1~
\mbox{~GeV}$ region. An analysis was carried out for pseudoscalar
and vector mesons with lepton decay constants, masses (obtained
from model dependent) and nucleon spin rules required to match
their experimental counterparts.

Possible behavior of  $\alpha_{s}$ with
$\alpha_{\mathrm{crit.}}=\alpha_{s}\left(Q^{2}=0\right) \sim
0.65-0.72$ in the case of a frozen regime, which follows from
experimental values of lepton decay constant, masses and nucleon
spin rules are discussed.
\end{abstract}

\section{Introduction \label{Sect1a}}

A running strong coupling constant $\alpha_{s}\left(Q^{2}\right)$
is one of the fundamental parameters of quantum chromodynamics. It
is of importance in many areas, such as non-relativistic  QCD,
description of quark-antiquark system, quark mass definitions and
others. Therefore, its behavior in the nonperturbative region
(small space-like momentum $Q < 1 \mbox{~GeV}$) is crucial for its
thorough description. Within a QCD framework, the behavior of
$\alpha_{s}\left(Q^{2}\right)$ is deduced from the solutions of
renormali\-za\-tion group equations.

In complete 4-loop approximation  the running coupling, obtained
in the $\mathrm{\overline{MS}}$-scheme, is given by
\cite{Chetyrkin:1997sg}:
\begin{eqnarray}
 \alpha_{\rm QCD}\left(Q^{2}\right)&=&
\pi\left[\frac{1}{\beta_0\; L_{Q}}-\frac{b_1\ln L_{Q}}{(\beta_0
L_{Q})^2}
  +\frac{1}{(\beta_0 L_{Q})^3}\left[b_1^2(\ln^2 L_{Q}-\ln
  L_{Q}-1)+b_2\right]+\right.
\nonumber\\
&&+\frac{1}{(\beta_0 L_{Q})^4}\left[
    b_1^3\left(-\ln^3 L_{Q}+\frac{5}{2}\ln^2 L_{Q}+2\ln
    L_{Q}-\frac{1}{2}\right)-\right.
    \nonumber\\
&&\left. \left. -3 b_1 b_2\ln L_{Q}+\frac{b_3}{2}\right]
\right]\;,\label{alhasqcd}
\end{eqnarray}
where we have used a shorthand notations
\begin{equation}
L_{Q} \equiv \ln z_Q=\ln(Q^2/\Lambda^2)\; , \; \; \; b_i =
\frac{\beta_i}{\beta_0}\ \; .\label{log}
\end{equation}
The $\beta$--functions are given by the equations
\begin{eqnarray}
&&\hspace{-10mm}
\beta_{0}=\frac{1}{4}\left(11-\frac{2}{3}n_{f}\right)\;,\;\;
\beta_{1}=\frac{1}{16}\left( 102 - \frac{38}{3} n_f\right)\;,
\nonumber\\
&&\hspace{-10mm}\beta_{2}=\frac{1}{64}\left(\frac{2857}{2} -
\frac{5033}{18} n_f + \frac{325}{54} n_f^2\right)\;, \nonumber\\
&&\hspace{-10mm} \beta_3 =\frac{1}{256}\left( \frac{149753}{6} +
3564 \zeta_3 + \left[- \frac{1078361}{162} - \frac{6508}{27}
\zeta_3 \right] n_f +\right.\nonumber
\\&&\mbox{}
  + \left[ \frac{50065}{162} + \frac{6472}{81} \zeta_3 \right] n_f^2
  +\left.  \frac{1093}{729}  n_f^3\right)\;, \label{betas}
\end{eqnarray}
where $n_f$ is the number of active flavours and $\zeta_{n}$ is
Riemann's zeta function.

The world average perturbative parameter is
\begin{equation}
\Lambda^{\left(nf=5\right)}_{\overline{MS}}=\left(231 \pm 8\right)
\mbox{~MeV}\; , \label{lambdaMS}
\end{equation}
which corresponds to \cite{Bethke2012}
\begin{equation}
\alpha_{\rm QCD}\left(M_Z^{2}\right)=0.1184 \pm 0.0007\; .
\label{alfamz}
\end{equation}

The presence of the Landau pole in (\ref{alhasqcd}) leads
$\alpha_{\rm QCD}$ to increase sharply at low $Q^2$. However,
there are numerous approaches, in which the behavior of the
coupling constant in the nonperturbative region differs
substantially from the conventional one (\ref{alhasqcd}). A number
of  models have been proposed for including nonperturbative
contributions at low $Q^2$:
\begin{enumerate}
\item Namely, instead of increasing indefinitely in the infrared, as
perturbation QCD predicts, it freezes at a finite value $\neq 0$
\cite{Richardson1978jl,Cornwall:1981zr,Godfrey:1985xj,Webber:1998um,Badalian:2001by,Shirkov:1997wi,Milton:2005hp,Shirkov:1999hm,Bakulev:2006ex,Shirkov2013mrg,Ganbold:2010bu,Aguilar:2009nf,Aguilar:2009vn}.
Most simply frozen mode constants can be obtained by replacing
(see, e.g.\cite{Shirkov2013mrg})
\begin{equation}
Q^2 \to Q^2+ m_g^2\; , \label{zamena}
\end{equation}
where  ``effective gluonic mass'' $m_g$ is some free parameter.

\item Models with the maximum in the nonperturbative
region, when $\alpha_{s} \to 0$ for $Q^ 2 \to 0$.
\cite{Dokshitzer:1995ev,Fischer:2008uz,Bornyakov:2013pha,Blossier:2011tf,Arbuzov:2007ae,Arbuzov:2013ofa}

\item Model, in which the growth constants is less than the QCD constant
\cite{Alekseev:1997rr,Alekseev:2004vx,Nesterenko:2003xb}.
\end{enumerate}

The freezing property  of the strong coupling constant at small
$Q^2$ is widely used in QCD-inspired hadron models
\cite{Richardson1978jl,Godfrey:1985xj,Badalian:1999fq,Peter:1997me}.
In the framework of the string model \cite{Kalashnikova:2001ig}
the QCD coupling is modifed so that it depends on the combination
$Q^2+m^2_{q}$ instead of $Q^2$ as it is in standard perturbative
theory. In two-loop approximation the running background coupling
is
\begin{eqnarray}
&&\hspace{-1.2cm} \alpha_{\rm BPT}^{(2)}\left(Q^{2}\right)=
\pi\left[\frac{1}{\beta_0\;L_{Q,m_g}}-\frac{b_1\ln
L_{Q,m_g}}{(\beta_0 L_{Q,m_g})^2}\right] \;,\;\;(
L_{Q,m_g}=\ln\left[\frac{Q^{2}+m_g^2}{\Lambda^{2}_{V}}\right])\;,
\label{alhasbpt}
\end{eqnarray}
where the mass $m_g \sim 1 \mbox{~GeV}$ is a background mass. It
was concluded \cite{Badalian:2001by}, that the most optimal
behavior of $\alpha_{s}$, is the one that leads to
\begin{equation}
\alpha_{\mathrm{crit.}}\equiv \alpha_{s}\left(Q^{2}=0\right) \sim
0.50 - 0.70 \; .\label{sq}
\end{equation}

Generalization of QCD coupling (\ref{alhasbpt}) can be obtained in
the framework of a ``massive'' perturbative renormalization group
(see, \cite{Shirkov:1999hm,Shirkov2013mrg} and references
therein).

An approximate two-loop  1-parameter model is of the form
\cite{Shirkov:1999hm,Shirkov2013mrg}:
\begin{equation}
\alpha_{\rm MPT}^{(2)}\left(Q^{2}\right)=\frac{\pi
\alpha_{\mathrm{crit.}}}{\pi+\alpha_{\mathrm{crit.}} \beta_0
\ln\left(1+z_Q/\xi \right)+b_1 \ln\left(1+ \alpha_{\mathrm{crit.}}
\beta_0 \ln\left(1+z_Q/\xi \right)/\pi \right)} \; ,\label{mpt}
\end{equation}
where parameter $\xi$ is expressed via the constant
$\alpha_{\mathrm{crit.}}$  by the relation
\begin{equation}
\xi=e^{\pi/\left(\alpha_{\mathrm{crit.}} \beta_0\right)}
\;.\label{mgl}
\end{equation}
It corresponds to the ``effective gluonic mass'' $m_g$
\begin{equation}
m_g=\sqrt{\xi} \Lambda\;.\label{mg}
\end{equation}

The analytic perturbation theory (APT) \cite{Shirkov:1997wi} (see
also Refs.
\cite{Milton:2005hp,Milton:2001pj,Shirkov:2006gv,Nesterenko:2003xb})
eliminates Landau pole. APT theory allows the property of
analyticity (and other ones) to be restored, which the standard
approach lacks. In the framework of the analytic approach instead
of (\ref{alhasqcd}), taken in the one-loop approximation, the
following expression was proposed to be used:
\begin{equation}
  \alpha_{\rm APT}^{(1)} (Q^{2}) = \frac{ \pi }{
   \beta_{0} } \left(
  \frac{1}{L_{Q}} +
  \frac{1}{1 -z_Q } \right)\;.
\label{shirsolconst}
\end{equation}
A crucial feature of the constant (\ref{shirsolconst}) is that for
$Q^{2} \to 0$ it takes a finite value, $\alpha_{{\mathrm{crit.}}}
\hm = \alpha_{\rm s}(0) = \pi / \beta_{0}\approx 1.4 \div 1.5 $,
and is independent of the renormalization scheme used, unlike
(\ref{alhasqcd}).

Based on analytic perturbation theory, the global fractional APT
was developed in \cite{Bakulev:2006ex}, in which the dependence
that  has $Q^2$ is different from that seen in (\ref{alhasqcd}).

Numerical solution of the nonperturbative effective coupling obtained in \cite{Cornwall:1981zr} is
given by
\begin{equation}
\alpha_{\rm Con}^{(1)} ( Q^{2} ) = \frac{\pi}{
   \beta_{0} } \left[
  \frac{1}{\ln(z_{Q}+4 M_g^{2}\left(Q^2\right)/\Lambda^{2})} \; \right]\;,
\label{conwallconst}
\end{equation}
where $M_g^{2}\left(Q^2\right)$ is dynamical gluon mass,
determined by the gluon mass $m_g$:
\begin{equation}
M_g^{2}\left(Q^2\right)=m_g^2 \left[ \frac{\ln \left(z_{Q}+4 m_g^{2}/\Lambda^{2}\right)}{\ln(4
m_g^{2}/\Lambda^{2})}\right]^{-12/11}\; . \label{conw1}
\end{equation}

In \cite{Webber:1998um}, the coupling constant
\begin{equation}
\alpha_{\rm W}^{(1)} ( Q^{2} ) = \frac{\pi}{
   \beta_{0} } \left[
  \frac{1}{L_Q} +
  \frac{1}{1-z_Q }\left(\frac{z_Q+ d}{1+d}\right)\left(\frac{1+c}{z_Q+c}\right)^{p} \; \right]\;
\label{weberconst}
\end{equation}
with parameters $d=1/4$ and $p=c=4$ is proposed for the estimation
of non-perturbative QCD power corrections.

The  freezing non-perturbative behavior of the QCD effective
charge $\alpha_{s}$, one obtained from the pinch technique gluon
self-energy, and one from the ghost-gluon vertex, is calculated in
\cite{Aguilar:2009nf,Aguilar:2009vn}. A fit for running constant
is provided by the following functional form
\begin{eqnarray}
&& \alpha_{\rm T}\left(Q^2\right)=\left[4\pi b \ln\left(\frac{Q^2+h(Q^2,m^2(Q^2))}{\Lambda^2}\right)\right]^{-1}\;, \nonumber \\
&& h(Q^2,m^2(Q^2))=\rho_{1} m^2(Q^2)+ \rho_{2} \frac{m^4(Q^2)}{Q^2+m^2(Q^2)}\;, \nonumber \\
&& m^2(Q^2)=\frac{m_0^4}{Q^2+ m_0^2} \frac{\ln\left(\left(Q^2+2
m_0^2\right)/\Lambda^2\right)}{\ln\left(2 m_0^2/\Lambda^2\right)}\;,\; \; \; b=33/(48 \pi^2)\;
,\label{alfat}
\end{eqnarray}
where $\rho_{1,2}$ and $m_0$ are fit parameters.

In \cite{Godfrey:1985xj} the behavior of effective strong coupling
constant is described by the phenomenological expression
\begin{equation}
\alpha_{\rm GI}(Q^2) = \sum_{k=1}^{3} \alpha_k \exp\left[-Q^2/
\left(4 \gamma^{2}_k\right) \right]\; \label{als}
\end{equation}
with coefficients  $\alpha_1 = 0.25,\;\alpha_2 = 0.15,\;\alpha_3 =
0.2$ è  $\gamma_1^2 = 1/4$, $\gamma_2^2 = 5/2$, $\gamma_3^2 =
250$.

In \cite{Dokshitzer:1995ev} a set of effective strong constants called $G_p$-models was offered to
explain experimental data on the hadronic jets initiated by heavy quarks:
\begin{eqnarray}
&& \alpha_{\rm D}^{(2)}\left(Q^{2}\right)= \pi\left(\frac{p \;
Q^{2 p}}{Q^{2p}+C_p \Lambda_{p}}\right)\left[\frac{1}{\beta_{0}
L_p}-\frac{b_{1} \ln L_p}{\left(\beta_{0} L_p\right)^{2}} \right
]\;,\label{alhasgm}
\end{eqnarray}
where
\begin{equation}\label{lp}
L_p=\frac{1}{p}\ln \left[ \frac{Q^{2 p}}{\Lambda^{2}}+C_p
\right]\;,C_p \geq 1\; , p=1,2, \ldots \;.
\end{equation}
The effective coupling (\ref{alhasgm}) has a maximum in the
nonperturbative region and $\alpha_{\rm D}^{(2)}\left(0\right)=0$.
A similar behavior has a running coupling constant calculated in
the framework of lattice gauge theory
\cite{Bornyakov:2013pha,Blossier:2011tf}.

The expression
\begin{equation}
\alpha_{\rm N}^{(1)} ( Q^{2} ) = \frac{\pi}{ \beta_{0} }
\left(\frac{z_Q-1}{ z_Q L_Q } \right)\;, \label{nesterconst}
\end{equation}
obtained from the requirement that the coupling constant should have correct analytic properties
(see, e.g., \cite{Nesterenko:2003xb}), leads to different behavior of the constant (\ref{alhasqcd})
in the region of small $Q^2$.

\begin{figure}[h t b]
\centering \vspace{0mm} \resizebox{0.7\textwidth}{!}{
\includegraphics{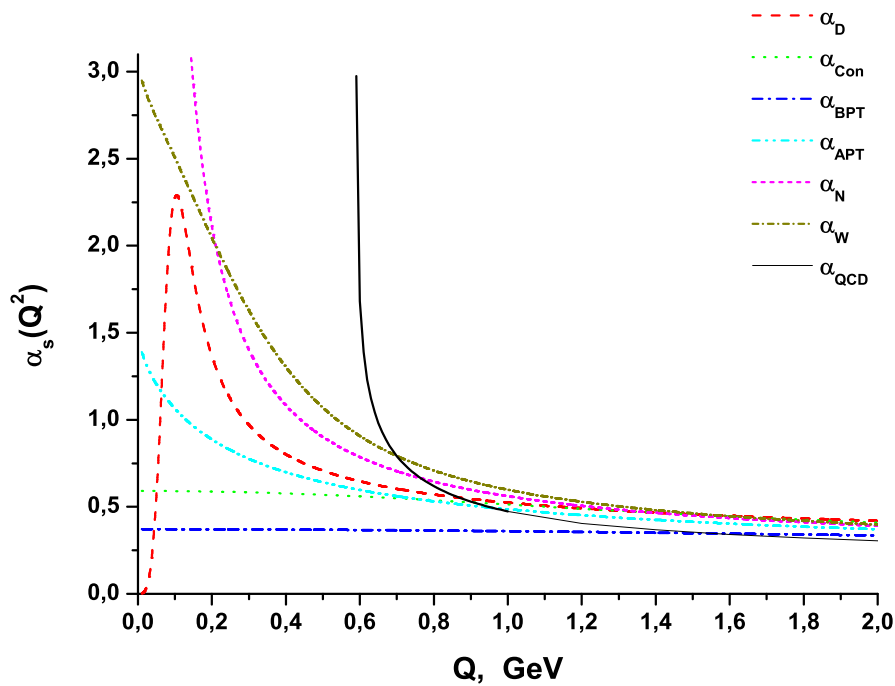}}
\vspace{0mm} \caption{{Various effective running strong coupling
constants (see Eqs. (\ref{alhasqcd}), (\ref{alhasbpt}),
 (\ref{shirsolconst}), (\ref{alhasgm}), (\ref{conwallconst})
(\ref{weberconst}), (\ref{nesterconst}))}} \label{qcdall}
\end{figure}

The main difference between all strong constants mentioned above
and the one quoted in (\ref{alhasqcd}) is that the effective
constants increase slower at small $Q^2$, whereas for $Q > 1
\mbox{~GeV}$ all these constants behave in a way nearly identical
to that of a standard QCD constant (Fig. \ref{qcdall}).

Thus, there is a multitude of models for the running coupling
constant as with a varying proportion between phenomenological and
theoretical motivations. That is why one of the primary goals in
this direction is to develop and improve methods that allow us to
determine the  QCD constant behavior.

Experimental determinations of $\alpha_{s}$ were regularly
summarised and reviewed in
\cite{Bethke:2009jm,BethkD:2006ac,Bethke2012}. These reviews
provide information on modern methods of obtaining values of the
constants from experimental data.

There are several techniques used to predict $\alpha_{s}$ at small
$Q^{2}$, e.g. the bound state approach which reproduces hadronic
characteristics and spectroscopy
\cite{Godfrey:1985xj,Badalian:1999fq,Baldicchi:2007zn,Baldicchi:2007ic},
lattice QCD \cite{Bornyakov:2009ug,Boucaud:2000ey}, solving the
Schwinger-Dyson equations
\cite{Cornwall:1981zr,Maris:1999nt,Aguilar:2009nf,Fischer:2008uz,Alkofer:2008dt},
pich technique \cite{Binosi:2009qm} and others. Matching the
results of theoretical calculations with experimental data should
lead to certain restrictions being put on the behavior of the
running constant, which is one of the model parameters.

In this work we to study the IR behavior of the constant using
combined approach, based on model calculations of meson
characteristics (bound state approach) and current computing of
pQCD corrections $ \hm \sim
\mathcal{O}\left(\alpha_{s}^{4}\right)$ to the sum rules of the
nucleon \cite{Baikov:2010je,Baikov:2012zn}. In this paper, we
develop the approach of Ref. \cite{Andreev2011ae}, which allows us
to investigate the possible behavior of QCD constant in the
infrared region.

To describe the properties of mesons we use Poincar\'{e}-covariant
quark model based on the principles of relativistic Hamiltonian
dynamics (RHD). The latter and their possible applications can be
found in \cite{Coester:1982vt,Keister:1991sb,Krutov2009rt}.

The basic requirement that restricts the possible behavior of
$\alpha_{s}\left(Q^{2}\right)$  in this method is a matching
condition between the model calculations and experimental values
of the leptonic decay constants, masses of pseudoscalar and vector
mesons and spin sum rules of nucleon.

The behavior of the modeling constant required to follow that of
the standard one (\ref{alhasqcd}) for $Q > 1 ~ \mbox{~GeV}$  is
considered an additional condition. Here, the behavior of the
running constant is simulated using improved phenomenological
parameterization (\ref{als})  for different sets of
$\alpha_{k},\gamma_{k}\;\left(k=1,\ldots,7\right)$. Using the
phenomenological constant (\ref{als}) significantly simplifies the
solution of the two-particle equation (\ref{maineq1}). Instead of
solving the equation (\ref{maineq1}) with different potentials,
which differ in the behavior of QCD constants, we solve one
equation with different sets of parameters $\alpha_{k}$ and
$\gamma_{k}$ (see Eq.(\ref{pvpotential})).

The layout of the paper is as follows. Section \ref{Sect2a}
contains information on the modeling of QCD constant behavior with
improved parameterization (\ref{als}).  A set of parameters that
simulate the behavior of $\alpha_{s}(Q^2)$ in the nonperturbative
region is obtained.

In Sec. \ref{Sect3a}--\ref{Sect4a}  Poincar\'{e}-covariant quark model of mesons  and calculation
of model leptonic meson decay constants are briefly described.

Sec. \ref{Sect5a}--\ref{Sect6a} is devoted to the strategy for
extracting ``optimal'' behavior of $\alpha_{s}$ from the
experimental data on leptonic constants and masses of mesons.

In Section \ref{Sect7a}--\ref{Sect9a} one can find an analysis of
the experimental and theoretical information about the nucleon sum
rules and the possible behavior of QCD constant in the infrared
region, which follows from this data.

\section{Modeling the effective coupling constant \label{Sect2a}}

To study infrared behavior of $\alpha_{s}$ one can try different
shapes of the effective coupling or, equivalently, different ways
to extrapolate the  improved para\-meteri\-zation (\ref{als}) for
different sets of
$\alpha_{k},\gamma_{k}\;\left(k=1,\ldots,7\right)$:
\begin{equation}
\alpha_{\mathrm{GI}}(Q^2) =  \sum_{k=1}^{n=7} \alpha_k
\exp\left[-Q^2/ \left(4 \gamma^{2}_k\right) \right]\;
\label{alsAn}
\end{equation}
to the IR region of small $Q^2$. Further, to identify a specific set of parameters, we use the
symbol $\texttt{N}_{\alpha}\; $.

This approach can be considered  QCD constant model independent
one, since we do not use any of the analytical expressions for the
constants (\ref{alhasqcd}), (\ref{alhasbpt}),
 (\ref{shirsolconst}), (\ref{alhasgm}), (\ref{conwallconst})
(\ref{weberconst}), (\ref{nesterconst}).

The behavior of the simulated constant required to follow that of
the standard one  (\ref{alhasqcd}) for $Q > 1-2 \mbox{~GeV}$ is
considered a necessary condition (within errors).

The values of the QCD constant (\ref{alhasqcd}) and corresponding
errors, which are used to calculate weighting coefficients, are
obtained using  \textsf{RunDec} program \cite{Chetyrkin:2000yt}.
The number of points to fit is varied from 550 to 600  over the
region of conformity (\ref{alsAn}) with the QCD constant. The
region of $Q$ varies from $0.6$ to $200$ GeV.

Since the restriction put on the behavior of the running constant
is based on the usage of a matching condition between experimental
and simulated values of the characteristics of pseudoscalar and
vector mesons, in which the constant  is integrated out, this
method will generally be ``sensitive'' to the square under the
curve, which shows as behavior.

For this reason it is unnecessary to use a function of type
(\ref{alhasqcd}); it is enough to get away with its approximation
(\ref{alsAn}), which must reproduce the well studied $Q
> 1.0 \mbox{~GeV}$ region.

We obtained sets of parameters differing in $\alpha_{\rm{crit.}}$
and the moment of the coupling (integral value):
\begin{equation}
{A}\left(\mu\right)=\frac{1}{\mu}\int\limits^{\mu}_{0}\mathrm{d}\mathrm{k}\;
\frac{\alpha_{s}\left(\mathrm{k}\right)}{\pi}\; \label{intDok}
\end{equation}
for $\mu=2~\mbox{~GeV}$,estimate of which was carried out in
\cite{Dokshitzer:1995ev,Dokshitzer:1999sh} (Tables \ref{tabfit},
\ref{tabfit1}).

Infrared behavior of the effective coupling constants can be
divided into two types: a) constant freezing with a smooth  and
monotonic increase
\cite{Cornwall:1981zr,Godfrey:1985xj,Shirkov:1997wi,Webber:1998um,Badalian:1999fq,Shirkov:1999hm}
and b) modes with the maximum, when $\alpha_{s} \to 0$ for $Q^ 2
\to 0$
\cite{Dokshitzer:1997ew,Bornyakov:2013pha,Blossier:2011tf,Aguilar:2009nf,Aguilar:2009vn}.

\begin{figure}[h t p]
\begin{tabular}{cc}
\subfigure[Freezing regime.]{\includegraphics[scale
=0.75]{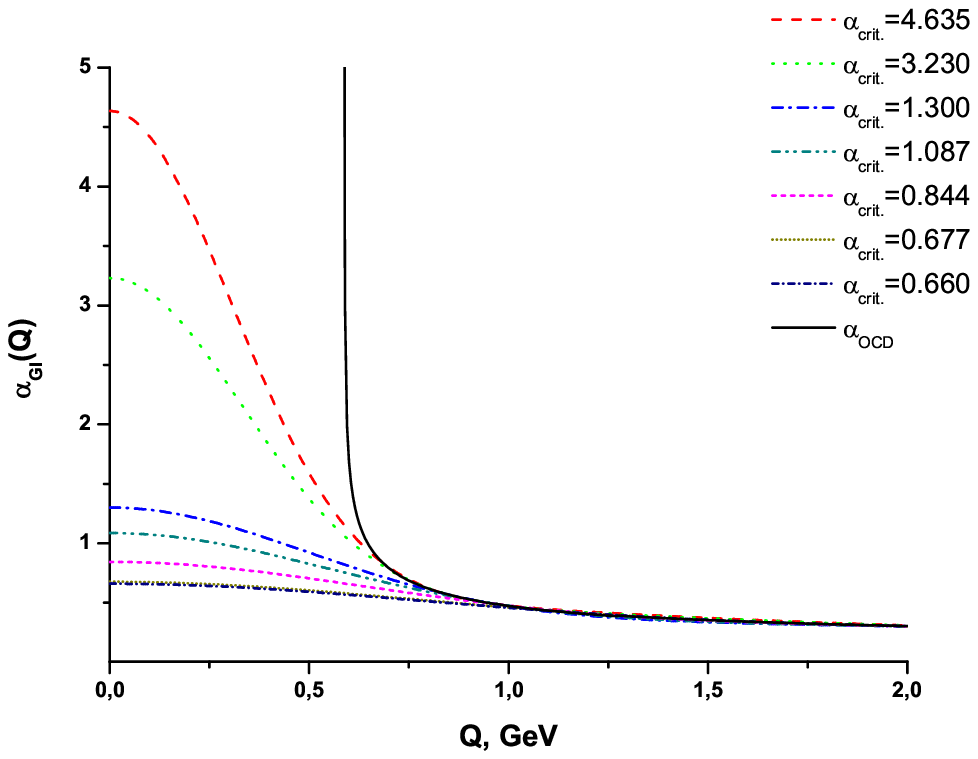}} &~
\subfigure[Modes with a peak in the nonperturbative
region.]{\includegraphics[scale =0.75]{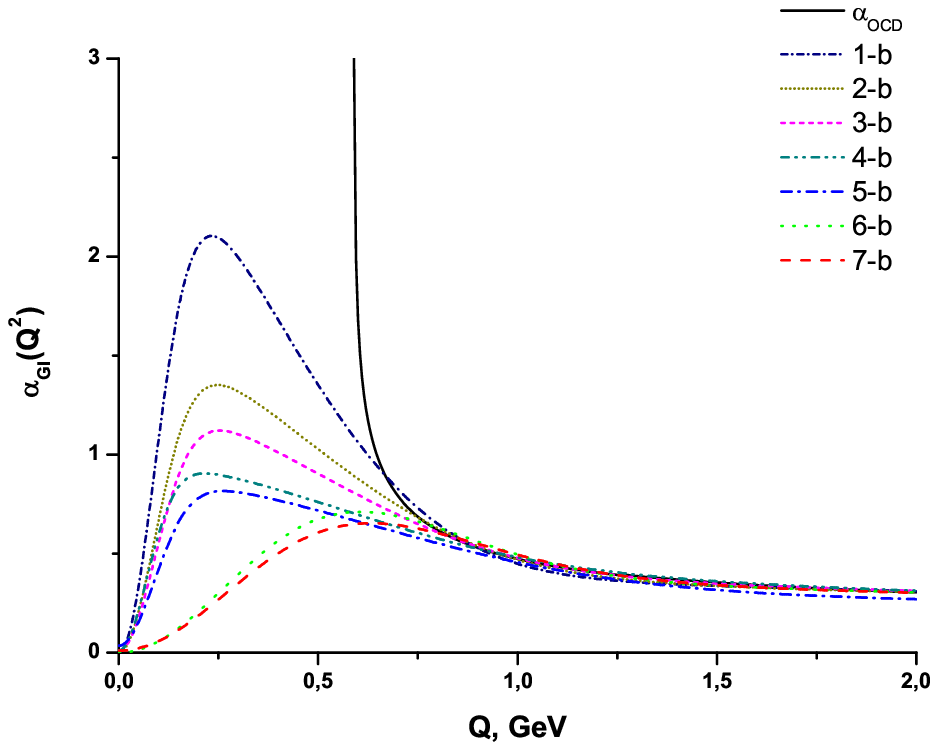}}
\end{tabular}
\caption{Dependence of the running strong coupling constant for
pQCD (\ref{alhasqcd}) and  pheno\-me\-nological parameterization
(\ref{alsAn}).\label{figalfas}}
\end{figure}

Therefore, we simulated both types of behavior: the first one
includes ``freezing'' the running coupling constant starting from
some value $Q_0 = 0.6 \div 1.0$ (see Table \ref{tabfit}), while
the second regime emulates the behavior with a peak in the
nonperturbative region (see Table \ref{tabfit1}).
\begin{table}[h t]
\begin{center}
\caption{\label{tabfit} Sets of constants (\ref{alsAn}) with
various $\alpha_{\mathrm{crit.}}$ for ``freezing'' regimes (see
left panel of Fig. ~\ref{figalfas}) .} \vspace{3mm}
\begin{tabular}{|c|c|c|}
  \hline
  $\texttt{N}_{\alpha}$, (number of set) & $\alpha_{\rm crit.}$ & $ {A}\left(2 ~\mbox{GeV}\right)$ \\
  \hline
  \texttt{1-a}  & $4.635\pm 0.006$  & $0.391  \pm 0.003  $\\
  \texttt{2-a}  & $3.230\pm 0.007$  & $0.318  \pm 0.003  $\\
  \texttt{3-a}  & $1.300\pm 0.003$  & $0.202  \pm 0.001  $\\
  \texttt{4-a}  & $1.087\pm 0.003$  & $0.186  \pm 0.001  $\\
  \texttt{5-a}  & $0.844\pm 0.003$  & $0.168  \pm 0.001  $\\
  \texttt{6-a}  & $0.687\pm 0.006$ & $0.155  \pm  0.002 $\\
  \texttt{7-a}  & $0.660\pm 0.007$  & $0.151  \pm 0.002  $\\
  \hline
\end{tabular}
\end{center}
\end{table}
\begin{table}[h  t ]
\begin{center}
\caption{\label{tabfit1} Sets of constants  (\ref{alsAn}) with
various $\alpha_{\mathrm{crit.}}$ for regimes with a peak in the
nonperturbative region (see right panel of Fig. ~\ref{figalfas}).}
\vspace{3mm}
\begin{tabular}{|c|c|}
  \hline
  $\texttt{N}_{\alpha}$, (number of set)  & $ \bar{A}\left(2 ~\mbox{GeV}\right)$ \\
  \hline
  \texttt{1-b}   & $0.238 \pm 0.014$\\
  \texttt{2-b}   & $0.192 \pm 0.008$\\
  \texttt{3-b}   & $0.178 \pm 0.016$\\
  \texttt{4-b}   & $0.165 \pm 0.009$\\
  \texttt{5-b}   & $0.150 \pm 0.005$\\
  \texttt{6-b}   & $0.134 \pm 0.023$\\
  \texttt{7-b}   & $0.127 \pm 0.015$\\
  \hline
\end{tabular}
\end{center}
\end{table}
A comparative graph for the coupling constants fixed by Eq.
(\ref{alhasqcd}) and various regimes of the effective constant
behavior (\ref{alsAn}) is demonstrated in Fig. \ref{figalfas}.

\section{Poincar\'{e}-covariant quark model of mesons \label{Sect3a}}

In our article we use the description of bound states with the
help of relativistic Hamiltonian dynamics (RHD)  that is the
generalization of the ordinary quantum mechanics.
\cite{Dirac:1949,Keister:1991sb}. RHD is also dubbed
Poincar\'{e}-invariant quantum mechanics (see, for instance,
\cite{Krutov2009rt}).

The RHD differs from the ordinary non-relativistic quantum
mechanics, as the main requirement for the operators of the
complete set of states is the one that the generators that make up
the operators should follow the algebra of Poincar\'{e} group.

In the framework of RHD, the interaction, which is determined by
the generators of the Poincar\'{e} group $\hat{P}_\mu $ and
$\hat{M}^{\mu \nu }$ is introduced as follows. The construction of
generators for a system of interacting particles starts from the
generators of an appropriate system composed out of noninteracting
particles, and then interaction is added so that the obtained
generators also satisfy the commutation relations of Poincar\'{e}
group. We shall not focus ourselves on the details of RHD and it's
connection with quantum field theory and special relativity but we
refer reader to the paper \cite{Keister:1991sb} and references
therein.

Unlike the case of a usual non-relativistic quantum mechanics, in
the relativistic case it is necessary to add interaction $\hat{V}$
in more than one generator to satisfy the algebra of the
Poincar\'{e} group. Dirac \cite{Dirac:1949} has shown that there
is no unambiguous separation of generators into dynamic set
(generators containing the interaction $\hat{V}$) and kinematic
set. There are three versions of separation on dynamic and
kinematic sets (so-called RHD forms):  point form, instant form
and dynamics on light front.

In all three forms the interaction contains mass operator
$\hat{M}$ i.e. $\hat{M}\equiv M_0+\hat{V}$, where $M_0$ is an
effective mass of a system of noninteracting  particles with
masses $m_q$ and $m_Q$:
\begin{equation}
M_0 = \sqrt{m_q^2+{\bf k}^2} + \sqrt{m_Q^2+{\bf k}^2} ~.
\label{va1}
\end{equation}
Here
\begin{eqnarray}
&&\hspace{-5mm} \mathbf{k} \mathbf{=}\frac{1}{2} \left(
\mathbf{p}_{1}-\mathbf{p}_{2}\right) +
\frac{\mathbf{P}}{\widetilde{M}_{0} \left(\omega
_{\widetilde{M}_{0}}\left( \mathrm{
P}\right) +\widetilde{M}_{0}\right)} \times \nonumber \\
&&\hspace{-5mm} \times \left(m_Q^2-m^2_q-\widetilde{M}_{0}\left[
\omega _{m_{Q}}\left( \mathrm{p}_{2}\right) -\omega _{m_{q}}\left(
\mathrm{p}_{1}\right) \right]\right) \label{va2}
\end{eqnarray}
is the relative momentum and $\mathbf{P}$ is the total momentum of
the free-system
\begin{equation}\label{va3}
\mathbf{P}=\mathbf{p_1+p_2}\; ,
\end{equation}
\begin{equation*}
\widetilde{M}_{0}= \sqrt{\left[\omega _{m_{Q}}\left(
\mathrm{p}_{2}\right) +\omega _{m_{q}}\left(
\mathrm{p}_{1}\right)\right]^2-\mathbf{P}^2}\label{va3as}
\end{equation*}
and $\omega_m\left(\mathrm{p}\right)=\sqrt{m^2+{\bf p}^2}$,
$\mathrm{k}=\left|\mathbf{k}\right|$.

In the framework of RHD the bound system with momentum
$\mathbf{Q}$, eigenvalues $M$, spin $J$ and it's projection $\mu $
is described by the wave function $ \Phi ^{J
\mu}_{\mathbf{Q};\;\sigma_{1} \sigma_{2}}\left({\bf k}\right)$ of
two-particle state, which satisfies the equation
\cite{Keister:1991sb}
\begin{eqnarray}
&&\hspace{-5mm} \sum_{\lambda_{1},\lambda_{2} }\int < {\bf
k},\sigma_{1},\sigma_{2}\parallel\hat{ V}\parallel \ {\bf
k}^{\prime}, \lambda_{1},\lambda_{2}> \times
\nonumber\\
&&\hspace{-5mm} \times \Phi ^{J \mu}_{\mathbf{Q};\lambda_{1}
\lambda_{2}}\left({\bf k}^{\prime }\right)\mathrm{d} {\bf
k}^{\prime} =\left(M-M_0\right) \Phi ^{J
\mu}_{\mathbf{Q};\sigma_{1} \sigma_{2}}\left({\bf k}\right)
\;.\label{maineq}
\end{eqnarray}
The radial equation for two-particle bound state in the
center-momen\-tum system ($\mathbf{Q}=0$) has the following form
\begin{equation}
\sum_{\ell^{\prime},S^{\prime}} \int\limits_{0}^{\infty}
V^{J}_{\ell,S\; ;\ell^{\prime},
S^{\prime}}\left(\mathrm{k},\mathrm{k^{\prime}}\right) \Phi^{J
\mu}_{\ell^{\prime}, S^{\prime}}\left(\mathrm{{k}^{\prime
}}\right){\mathrm{{k}^{\prime}}}^{2}\mathrm{d}{\mathrm{k}}^{\prime}=
\left(M- M_0\right) \Phi ^{J \mu}_{\ell,
S}\left(\mathrm{{k}}\right)\;. \label{maineq1}
\end{equation}

To describe specific bound systems, it is necessary to determine
the interaction potential between particles. It should be noted
that different potentials can be used to describe a bound system
of the same composition. Such a selection of potentials
automatically distinguishes different Poincar\'{e}-covariant
models.

In our case the interquark potential in coordinate representation
from \cite{Godfrey:1985xj} is used, which is considered a sum of
Coulomb, linear confinement, and spin-spin parts for pseudoscalar
and vector mesons:
\begin{eqnarray}
&&\hat V\left(\mathrm{r}\right) = \hat{V}_{Coulomb}(\mathrm{r})+
\hat{V}_{linear}(\mathrm{r})+ \hat{V}_{{S}{S}}(\mathrm{r}) \;,  \label{pvpotential} \\
&&\hat{V}_{Coulomb}(\mathrm{r})=- \frac{4}{3}\;
\frac{\alpha_{s}\left(\mathrm{r}\right)
}{\mathrm{r}}=-\frac{4}{3\;\mathrm{r}}\; \sum_{k=1}^7 \alpha_k
\,\hbox{erf}(\tau_k\,\mathrm{r})\;,
\nonumber \\
&&\hat{V}_{linear}(\mathrm{r})=\sigma\;
\mathrm{r}\left[\frac{\exp(-b^{2} \mathrm{r}^{2})} {\sqrt{\pi}\;b
\;\mathrm{r}}+ \left(1 + \frac{1}{2\; b^2 \mathrm{r}^2}\right)\;
\hbox{erf}(b\;\mathrm{r})\right] + w_0\;,
\nonumber \\
&&\hat{V}_{{S}{S}}(\mathrm{r})= -\frac{32\;\left( {\mathbf{S}}_{q}
{\mathbf{S}}_{Q}\right)}{9 \sqrt{\pi}\; m_q\; m_Q}\sum_{k=1}^7
\alpha_k\; \tau_k^3\;\exp(-\tau_k^2 \mathrm{r}^2)\;,\nonumber
\end{eqnarray}
where parameter $\tau_k$ is deduced from the relation
${1/\tau_k^2} = {1/\gamma_k^2} + {1/\mathrm{b}^2}$,
$\hbox{erf}(x)$) is an error function, and $\mathbf{S}_{q,\;Q}$
denote quark spin operators.

To derive the potential (\ref{pvpotential}) given in, the procedure of ``smearing'' was applied
according to the following rule \cite{Godfrey:1985xj}:
\begin{equation*}\label{razmaz}
\tilde{f}\left(\mathbf{r}\right)=\int
\mathrm{d}\mathbf{r}^{\prime}
\rho\left(\mathbf{r}-\mathbf{r}^{\prime}\right)f\left(\mathbf{r}^{\prime}\right)
\;,
\end{equation*}
where the ``smearing'' function with parameter $\mathrm{b}$  is
chosen in the form
\begin{equation*}\label{smear}
\rho\left(\mathbf{r}-\mathbf{r}^{\prime}\right)=\frac{\mathrm{b}^{3}}{\pi^{3/2}}
\exp\left[-\mathrm{b}
\left(\mathbf{r}-\mathbf{r}^{\prime}\right)^{2}\right]\;.
\end{equation*}

\section{Leptonic meson decay constants in Poincar\'{e}-covariant model \label{Sect4a}}

Upon the removal of element $V_{Q q}$ of the
Cabibbo-Kobayashi-Maskawa matrix, the constant  $f_P$ of the
leptonic decay  $f_P$$ P (Q\bar q)\to \ell+\nu_{\ell}$ for a
pseudoscalar meson $P (Q\bar q)$ is  defined by the relation:
\begin{equation}
\label{dec2} j^\mu_P \equiv \left < 0\left| \hat {J}^{\mu}_{A}
\left( 0\right) \right| {\bf P},M_P\right
>_{in} =\mathrm{i} \left( 1/2\pi \right) ^{3/2}\frac{P^\mu f_P}{\sqrt{2\;\omega _{M_P}\left(
{\mathrm{P}}\right) }}  \;,
\end{equation}
where the electroweak axial current $\hat {J}^{\mu}_{A}(0)$ and
the vector of a meson state with mass $M_P$ are taken in the
Heisenberg representation. Vectors of states in this expression
are normalized as follows: $\left<  \mathbf{P},M_{P} \right.\left|
\mathbf{P}^{\prime},M_{P}\right>$=$\delta( \mathbf{P}-
\mathbf{P}^{\prime})$.

The decay width for  $P (Q\bar q)\to \ell+\nu_{\ell}$ is given by
the expression
\begin{eqnarray}
&& \Gamma_{P}=\frac{G_{F}^{2}\left|V_{Q q}\right|^{2}
}{8\pi}m_{\ell}^{2}M_{P} f_{P}^2
\left(1-\frac{m_{\ell}^{2}}{M_P^2}\right)^{2} \;, \label{widthP}
\end{eqnarray}
where $m_{\ell}$ is lepton mass and $G_{F}$ is a Fermi constant.

In the case of leptonic decays of vector mesons $V(Q\bar q)\to
\ell+ \bar{\ell}$ relations analogous to the expressions given in
(\ref{dec2}) and (\ref{widthP}) will take the form
\begin{equation*}
\label{dec2V} j^\mu_V \equiv \left < 0\left| \hat {J}^{\mu}_{V}
\left(0\right) \right| {\bf P},M_V,\lambda\right>_{in} =\mathrm{i}
\left( 1/2\pi \right) ^{3/2} \frac{\varepsilon^\mu_{\lambda} M_V
f_V}{\sqrt{2\;\omega _{M_V} \left( {\mathrm{P}}\right) }}
\end{equation*}
with $\varepsilon^\mu_{\lambda}$  being the polarization vector
of a vector meson with mass $M_V$. Respectively, the decay width
for $V(Q\bar q)\to \ell+\bar{\ell}$ is given by the following
expression:
\begin{eqnarray}
&& \Gamma_{V}=\frac{4 \pi \alpha^{2}}{3\;M_{V}}f_{V}^2
\left(1+\frac{2\;m_{\ell}^{2}}{M_V^2}\right)\sqrt{1-\frac{4\;m_{\ell}^{2}}{M_V^2}}
\;, \label{widthV}
\end{eqnarray}
where $\alpha$ stands for the fine structure constant. In
 \cite{Andreev2000ae,Krutov:1997wu}  coinciding integral representations for
the constants of leptonic pseudoscalar $f_P$ and vector $f_V$,
meson decays are obtained within the framework of
Poincar\'{e}-covariant models based on the point and instant forms
of RHD:
\begin{eqnarray}
&&\hspace{-1.2cm} f_P\left(m_q,m_Q\right)= \frac{N_c}{\pi
\sqrt{2}}\int\limits_0^\infty \mathrm{d}\mathrm{k}\hskip 2pt
\mathrm{k}^2 \psi^P \left(\mathrm{k}\right)
\sqrt{\frac{M_0^2-(m_q-m_Q)^2}{\omega _{m_q}\left( \mathrm{k}
\right) \omega _{m_Q}\left( \mathrm{k}\right)}}\frac{\left(
m_q+m_Q\right) }{M_0^{3/2}}\;, \label{leptonconst}
\end{eqnarray}
\begin{eqnarray}
&&f_{V}\left( m_{q},m_{Q}\right)=\frac{N_{c}}{\sqrt{2}\pi }
\int\limits_{0}^{\infty }\mathrm{d}\mathrm{k}\hskip 2pt
\mathrm{k}^2 \psi^{V}\left( \mathrm{k}\right) \frac{\sqrt{\left(
\omega _{m_{q}}\left( \mathrm{k}\right) +m_{q}\right) \left(
\omega _{m_{Q}}\left( \mathrm{k}\right) +m_{Q}\right)
}}{\sqrt{\omega _{m_{q}}\left( \mathrm{k}\right) +\omega
_{m_{Q}}\left( \mathrm{k}\right) }\; \omega _{m_{q}}\left(
\mathrm{k}\right) \omega _{m_{Q}}\left( \mathrm{k}
\right) }\times   \nonumber \\
&&\times \left( 1+\frac{\mathrm{k}^{2}}{3\left( \omega
_{m_{q}}\left( \mathrm{k}\right) +m_{q}\right) \left( \omega
_{m_{Q}}\left( \mathrm{k}\right) +m_{Q}\right) }\right) \;,
\label{fvconst}
\end{eqnarray}
where $N_c$ is a number of quark colors.

Analogous integral representation for $f_P$ is derived in
\cite{Jaus:1991cy} in the context of the Poincar\'{e}-covariant
model based on the light front dynamics. The representations in
(\ref{leptonconst}) and (\ref{fvconst}) in the nonrelativistic
case become classical expressions in which constants are directly
proportional to the meson wave function in the position
representation at the origin $\mathrm{r}=0$.

\section{Selection of model parameters \label{Sect5a}}

Let us solve the eigenvalue problem (\ref{maineq1}) with potential (\ref{pvpotential}) by using
variational method with oscillator and Coulomb (for $B$-mesons) wave functions. In this method it
is required to minimize the functional
\begin{eqnarray*}
&& M\left( m_{q},m_{Q},\beta ,w_{0},b,\sigma \right) =  \langle
\psi\left(\beta\right) |{\hat M}|\psi\left(\beta\right)  \rangle =
\langle \psi|\hat M_0|\psi \rangle + \langle \psi|\hat V|\psi
\rangle\;, \label{MP}
\end{eqnarray*}
where $\psi\left(\beta\right)$ is a trial wave function.

The potential of the model (\ref{pvpotential}) has the following
free parameters: gluon string tension $\sigma$, smearing factor
$\mathrm{b}$, and $w_0 $. Quark masses $m_{q,Q}$ and sets of
constants $\alpha_{k},\gamma_{k}$, which characterize the behavior
of the effective strong coupling constant, are also considered as
parameters. Note that the values of parameters $\beta, w_0,
\sigma$ depend on the quark flavors.

Let's consider a procedure for fixing the numeric values of the
potential parameters. The parameter of the potential's linear part
lies within the range of $0.18$ to $0.20$ $\mbox{GeV}^2$
\cite{Godfrey:1985xj,Ebert:1999xv,Balandina:2000gz,Kalashnikova:2001ig}
in a large number of models. Therefore, we assume in the
calculations that
\begin{equation}\label{parstring}
\sigma = \bar{\sigma} \pm \Delta\sigma \;=\; \left(0{.}19 \pm
0.01\right) \;\;\mbox{GeV}^2\;.
\end{equation}

Wave function parameter  $\beta$ and all other potential
parameters are determined by solving the following system of
equations:
\begin{eqnarray}
&&\hspace{-1.2cm} \frac{\partial M_{P,V}\left(\beta, \sigma
\right)}{\partial \beta}\left.\right.
\bigg|_{\beta_{min},\tilde{\sigma}}=0~\;,\;\;M_{P}\left(w_0,\beta_{min},\tilde{\sigma}\right)=M_P
\pm \Delta M_P\;,
\label{system}\\
&&\hspace{-1.2cm} M^{S=1}_{V}\left(\beta,\sigma\right)-
M^{S=0}_{P}\left(\beta,\sigma\right) \left.\right|_{\beta_{min},
\tilde{\sigma}}=M_V-M_P \pm \delta M_{\rm{vp}} \;,
\label{system1}\\
&& \hspace{-1.2cm} f_{P}\left(m_q, m_Q,
\beta_{min}\right)=f_{\rm{exp}}^{P} \pm \Delta
f_{\rm{exp}}^{P}\;,\label{system3}\\
&& \hspace{-1.2cm} f_{V}\left(m_q, m_Q,
\beta_{min}\right)=f_{\rm{exp}}^{V} \pm \Delta
f_{\rm{exp}}^{V}\;,\label{system4}
\end{eqnarray}
where Eqs. (\ref{system}) and (\ref{system1}) are minimum
condition and requirement for simulated values of meson masses to
match their experimental counterparts. Quantities $M_{P,V}$ are
experimental values of pseudoscalar and vector meson masses and
$\Delta M_{P,V}$ are their experimental measurement errors. The
last two equations mean that the values of leptonic coupling
constants for pseudoscalar and vector mesons, obtained using the
Poincar\'{e}-covariant model coincide (see (\ref{leptonconst}),
(\ref{fvconst})) with experimental values $f_{\rm{exp}}$ within
the errors.

\subsection{Masses of $u,d$ and $s$ quarks }
Assuming that constituent masses of u and d quarks are
approximately equal \cite{Godfrey:1985xj}:
\begin{equation}
 m_d-m_u \equiv \Delta m_{ud} =\left(4 \pm 1\right)\mbox{MeV}\; , \label{udm}
\end{equation}
we obtain a system of equations from
(\ref{system})--(\ref{system4}):
\begin{equation*}
\left\{
\begin{array}{l}
f_{V}\left(m_u, m_d, \beta \right)=f_{\rm{exp}}^{\rho^{0}} \pm
\Delta f_{\rm{exp}}^{\rho^{0}}\;,\\
f_{P}\left(m_u, m_d, \beta \right)=f_{\rm{exp}}^{\pi^{\pm}} \pm
\Delta f_{\rm{exp}}^{\pi^{\pm}}\;.\\
\end{array}
\right. \label{systpion0}
\end{equation*}

Using experimental data for $\pi^{\pm}$ and $\rho^{0}$ mesons
\cite{Rosner:2008yu}
\begin{eqnarray*}
&&\hspace{-1.2cm} f_{\pi^{\pm}}^{P}=\left(130.41 \pm 0.03 \pm
0.20\right)~ \mbox{MeV}\;,\;\; f_{\rho^{0}}^{V}=\left(156.2 \pm
1.2\right)~ \mbox{MeV}\;, \label{expr}
\end{eqnarray*}
where the last numerical value is obtained from (\ref{widthV}) and
the experimental data of width $\Gamma_{\rho^{0}}= 7.02 \pm 0.11
\mbox{~~keV}$ for the decay $\rho^{0} \to e^{+}e^{-}$
\cite{pdg2012}, we obtain the values of $u$- and $d$-quark masses:
\begin{equation}
m_u = \left(239.8 \pm 2.3\right)\mbox{MeV}\;\; ,\; \;m_d =
\left(243.8 \pm 2.3\right)\mbox{MeV}\;.
\label{udmacca}
\end{equation}

Depending on the behavior of the running strong coupling constant
$\alpha_{s}\left(Q^{2}\right)$, the solution of equations system
\begin{equation}
\left\{
\begin{array}{l}
\partial M_{V}\left(\beta, \ldots\right)/\partial \beta=0~\;,\;\;
M_{K^{+}}\left(\beta, \ldots\right)=M_{K^{\pm}} \pm \Delta M_{K^{\pm}}\;,\\
M^{S=1}_{V}\left(\beta, \ldots\right)- M^{S=0}_{P}\left(\beta,
\ldots \right)=M_{K^{*}}-M_{K^{\pm}} \pm \delta M_{K^{*}-K^{\pm}}\;, \\
f_{P}\left(m_u, m_s, \beta \right)=f_{\rm{exp}}^{K^{\pm}} \pm
\Delta
f_{\rm{exp}}^{K^{\pm}}\;.\\
\end{array}
\right. \label{systpionplus}
\end{equation}
with account for the experimental data \cite{pdg2012}
\begin{eqnarray*}
&&\hspace{-1.2cm} M_{K^{+}}=\left(493.677 \pm 0.016\right)
~\mbox{MeV}\;,\;\;f_{K^{\pm}}^{P} = \left(156.1 \pm 0.2 \pm 0.8
\pm 0.2\right) ~ \mbox{MeV}\;,
\nonumber\\
&& \Delta M_{\mathrm{exp}}=M_{K^{*}}-M_{K^{\pm}}=\left(397.983~\pm
0.261 \right) \mbox{~~MeV}\; \label{expr2}
\end{eqnarray*}
and the value of $u$-quark mass (\ref{udmacca}) gives the results
presented in Table \ref{massudtab} with experimental and
theoretical uncertainties indicated.
\begin{table}[b]
\begin{center}
\caption{ Allowed values of  $s$--quark mass for different regimes
$\texttt{N}_{\alpha}$ of $\alpha_{s}$-behavior (with various
$\alpha_{\mathrm{crit.}}$ and ${A}\left(\mu\right)$
(\ref{intDok})). \label{massudtab}} \vspace{2mm}
\begin{tabular}{|c|c|c||c|c|}
  \hline
$\texttt{N}_{\alpha}$ & $\alpha_{\rm crit.} $ & $m_s\;,\mbox{~MeV}$ & $\texttt{N}_{\alpha}$ & $m_s\;,\mbox{~MeV}$\\
  \hline
  \texttt{1-a} & $4.635\pm 0.006$ & $478.4  \pm 23.1$ &\texttt{1-b} &  $467.2  \pm 28.9$ \\
  \texttt{2-a} & $3.230\pm 0.007$ & $469.1  \pm 23.5$ &\texttt{2-b} &  $461.6  \pm 28.7$ \\
  \texttt{3-a} & $1.300\pm 0.003$ & $460.4  \pm 25.0$ &\texttt{3-b} &  $461.5  \pm 31.1$ \\
  \texttt{4-a} & $1.087\pm 0.003$ & $461.0  \pm 25.6$ &\texttt{4-b} &  $463.0  \pm 30.6$ \\
  \texttt{5-a} & $0.844\pm 0.003$ & $463.3  \pm 26.7$ &\texttt{5-b} &  $460.9  \pm 29.9$ \\
  \texttt{6-a} & $0.687\pm 0.006$ & $466.5  \pm 28.0$ &\texttt{6-b} &  $471.1  \pm 29.9$ \\
  \texttt{7-a} & $0.660\pm 0.007$ & $466.6  \pm 28.0$ &\texttt{7-b} &  $473.2  \pm 30.7$ \\
  \hline
\end{tabular}
\end{center}
\end{table}
\subsection{Masses of $c$ and $b$ quarks}

To calculate leptonic constants for heavy mesons we need to know
the masses of $c$ and $b$ quarks. In order to compute the
constraints on the values of quarks the data on $c\bar{c}$
($\eta_{c}$ and $J/\psi$ mesons) and $b\bar{b}$ ($\eta_{b}$ and
$\gamma \left(1S\right)$ mesons) systems are used:
\begin{eqnarray}
&&\hspace{-1.2cm} M_{\eta_{ñ}}=\left(2981.0 \pm 1.1\right)
~\mbox{MeV}\;, \;\; M_{J/\psi}=\left(3096.916 \pm  0.011\right)
~\mbox{MeV}\;, \nonumber\\
&& \hspace{-1.2cm} M_{\eta_{b}}=\left(9391.0 \pm 2.8\right)
~\mbox{MeV}\;, M_{\gamma \left(1S\right)}=\left(9460.30 \pm
0.26\right) ~\mbox{MeV}\;. \label{ccbbexpr}
\end{eqnarray}

Since these systems consist of particles with  equal masses, it is
enough to use experimental data only for leptonic decays of vector
states in order to fix the masses of the quarks:
\begin{eqnarray}
&& \hspace{-1.2cm} f_{\gamma \left(1S\right)}^{V}=  \left(238.4
\pm  1.6\right) ~ \mbox{MeV}\;,\;\; f_{J/\psi}^{V}=\left(277.6 \pm
4\right)  ~\mbox{MeV}\;. \label{fvccbbexpr}
\end{eqnarray}

A solution to the system of equations analogous to Eqs.
(\ref{systpionplus}) leads to constraints on the masses of $c$ and
$b$ quarks that are presented in Table ~\ref{masscbtab}.
\begin{table}
\begin{center}
\caption{Allowed values of $c$ and $b$ quarks masses for different
regimes $\texttt{N}_{\alpha}$ of the $\alpha_{s}$-behavior (with
various $\alpha_{\mathrm{crit.}}$ and ${A}\left(\mu\right)$
(\ref{intDok})). \label{masscbtab}} \vspace{2mm}
\begin{tabular}{|c|c|c||c|c|c|}
  \hline
  $\texttt{N}_{\alpha}$ & $m_c\;,\mbox{~GeV} $ & $m_b\;,\mbox{~GeV}$ & $\texttt{N}_{\alpha}$ & $m_c\;,\mbox{~GeV}$ & $m_b\;,\mbox{~GeV}$\\
  \hline
  \texttt{1-a}  & $1.500 \pm 0.068$ & $3.668 \pm 0.158$ &\texttt{1-b}  & $1.454 \pm 0.080$ & $3.814  \pm 0.121 $ \\
  \texttt{2-a}  & $1.473 \pm 0.069$ & $3.748 \pm 0.159$ &\texttt{2-b}  & $1.420 \pm 0.077$ & $3.919  \pm 0.130 $ \\
  \texttt{3-a}  & $1.410 \pm 0.068$ & $3.965 \pm 0.174$ &\texttt{3-b}  & $1.410 \pm 0.091$ & $3.746  \pm 0.258 $ \\
  \texttt{4-a}  & $1.399 \pm 0.069$ & $3.995 \pm 0.174$ &\texttt{4-b}  & $1.396 \pm 0.082$ & $3.965  \pm 0.126 $ \\
  \texttt{5-a}  & $1.385 \pm 0.069$ & $4.029 \pm 0.178$ &\texttt{5-b}  & $1.372 \pm 0.074$ & $4.281  \pm 0.228 $ \\
  \texttt{6-a}  & $1.373 \pm 0.071$ & $4.057 \pm 0.157$ &\texttt{6-b}  & $1.382 \pm 0.100$ & $4.034  \pm 0.059 $ \\
  \texttt{7-a}  & $1.366 \pm 0.070$ & $4.092 \pm 0.180$ &\texttt{7-b}  & $1.373 \pm 0.091$ & $4.066  \pm 0.090 $ \\
  \hline
\end{tabular}
\end{center}
\end{table}

\section{Determination of the ``optimal''  $\texttt{N}_{\alpha}$ \label{Sect6a}}

Optimal value for $\alpha_{\rm{crit.}}$ and, respectively, a
possible regime of $\alpha_{s}\left(Q^2\right)$ behavior will be
chosen by using the experimental data on the constants of leptonic
decays of pseudoscalar heavy mesons ($D$ and $D_s$ mesons).

The quantity $\chi^{2}$ will be the main criterion used in this
selection procedure. To this end, let us compute the $\chi^{2}$
quantity
\begin{equation}
\label{chi2} \chi^{2}\left(\texttt{N}_{\alpha}\right)= \sum_{i=
1}^{k} \frac{\left( f^{P}_{i,\;
\mathrm{exp}}-f^{P}_{i}\left(m_q,m_Q,\beta \right)\right)^{2}}{
\left(\delta f^{P}_{i,\;\mathrm{exp}}\right)^{2}+ \left(\delta
f^{P}_{i,\;\mathrm{teor}}\right)^{2}}
\end{equation}
for various regimes of  strong coupling constant behavior and find
its minimum. The quantity $\delta f^{P}_{\mathrm{teor}}$ includes
all uncertainties related with both theoretical (originating from
calculations) and experimental errors for meson masses whose
values served as benchmarks for finding model parameters. In the
asymptotic limit, the quantity $
\chi^{2}\left(\texttt{N}_{\alpha}\right)$ will be distributed like
$\chi^{2}$ with $k$ degrees of freedom.

Using the following experimental values from \cite{Rosner:2008yu}
\begin{eqnarray*}
&&\hspace{-1.2cm} f_{D,\mathrm{exp}}^{P}=(206.7 \pm 8.9 )~
\mbox{MeV}\;,\;f_{D_s,\mathrm{exp}}^{P}=(260.0 \pm 5.4) ~\mbox{MeV}
\label{fpexprbdmeson}
\end{eqnarray*}
and calculations of the leptonic constants for pseudoscalar mesons
within a framework of the Poincar\'{e}-covariant model, we find
the dependence of $\chi^{2}\left(\texttt{N}_{\alpha}\right)$ has
on the regimes of $\alpha_{s}$ behavior. These results we
presented in Table \ref{chi2alfa}, along with the acceptance
probabilities $P$ (in $\%$).
\begin{table}
\begin{center}
\caption{ The values of $\chi^{2}\left(\texttt{N}_{\alpha}\right)$
(\ref{chi2}) and acceptance probabilities $P$ of model for various
regimes of $\alpha_{s}$ behavior. \label{chi2alfa}} \vspace{2mm}
\begin{tabular}{|c|c|c||c|c|c|}
  \hline
  $\texttt{N}_{\alpha}$ & $\chi^{2}\left(\texttt{N}_{\alpha}\right)$ & $P,\;\%$ &
  $\texttt{N}_{\alpha}$ & $\chi^{2}\left(\texttt{N}_{\alpha}\right)$ & $P,\;\%$\\
  \hline
  \texttt{1-a} & $6.74  $ & $3.4 $ &\texttt{1-b}  & $3.07 $ & $21.5$ \\
  \texttt{2-a} & $4.63  $ & $9.9 $ &\texttt{2-b}  & $1.51 $ & $46.9$ \\
  \texttt{3-a} & $1.33  $ & $51.3$ &\texttt{3-b}  & $0.77 $ & $68.1$ \\
  \texttt{4-a} & $0.84  $ & $65.8$ &\texttt{4-b}  & $0.39 $ & $82.1$ \\
  \texttt{5-a} & $0.32  $ & $85.0$ &\texttt{5-b}  & $0.34 $ & $84.2$ \\
  \texttt{6-a} & $0.08  $ & $96.1$ &\texttt{6-b}  & $0.06 $ & $97.2$ \\
  \texttt{7-a} & $0.05  $ & $97.6$ &\texttt{7-b}  & $0.02 $ & $99.0$ \\
  \hline
\end{tabular}
\end{center}
\end{table}

As follows from the calculations (see, Table \ref{chi2alfa})
models with freezing constant $\texttt{N}_{\alpha}\hm
=\texttt{6-{a}}, \texttt{7-{a}}$ have a minimum of $\chi^{2}$ and
the maximum acceptance probability  $P > 95\; \%$ . The highest
acceptance probability ($P>95\; \%$) for modes with a peak are
constants for the sets $\texttt{N}_{\alpha}=\texttt{6-{b}}$ and
$\texttt{N}_{\alpha}=\texttt{7-{b}}$.

Model leptonic constants of  $D$ and $D_s$ mesons for $\texttt{N}_{\alpha}=\texttt{7-{a}}$ are
\begin{eqnarray*}
&&\hspace{-1.2cm} f_{D}^{P}=(204.6  \pm  4.5 )~ \mbox{MeV}\;,\;f_{D_s}^{P}=(259.3 \pm 10.4)
~\mbox{MeV}\; .\label{fmodel}
\end{eqnarray*}
Values  (\ref{fmodel}) are in good agreement with the data
(\ref{fpexprbdmeson}).

However, it should be noted that models \texttt{5-\rm{a}},
\texttt{6-{a}}, and \texttt{4-{b}}, \texttt{5-{b}}, as well as
\texttt{6-{b}} have relatively larger probabilities and cannot be
definitely discarded since their $\chi^{2}/2 \leq 1$. For further
restrictions additional information is needed. The data on
$B$-meson decays can provide such information.

At present, from our point of view, there is a considerable
variation in determining the leptonic constant of a charged $B$
meson. An experimental value of the quantity (see
\cite{Aubert:2007xj,Aubert:2007bx,Adachi:2008ch,Schwartz:2009hv,Aubert:2009rk})
\begin{equation*}\label{bmesexp}
f_{B}^P \left|V_{ub}\right|=\left(7.2 \div 10.1\right)\times
10^{-1}~\mbox{MeV}\;,
\end{equation*}
with a modern constraint imposed on
$\left|V_{ub}\right|=\left(4.15 \pm 0.49\right)\times 10^{-3}$
\cite{Kowalewski2012} entails substantial variation in $f_{B}^P$:
\begin{equation*}\label{fbexp}
f_{B}^P=\left(208.4 \pm 42.7\right)~\mbox{MeV}\;.
\end{equation*}
There also is a considerable variation in theoretical predictions
for the leptonic decay constants of ${B}^{\pm}$-mesons: from
$f_{B}^P=\left(147^{+34}_{-38} \right)~\mbox{MeV}$  in
\cite{AliKhan:1998df} to $f_{B}^P \hm =\left(230 \pm 23\right)
~\mbox{MeV}$  in \cite{Penin:2001ux}.

In our approach, for the optimal regime with freezing constant
\texttt{7-\rm{a}}, the leptonic decay constant is found to be
\begin{equation}\label{fbexp1}
f_{B}^P=\left(226.2 \pm 3.9 \right)~\mbox{MeV}\;,
\end{equation}
while for the regime with a peak \texttt{7-{b}} we have
\begin{equation}\label{fbexp2}
f_{B}^P=\left(216.7 \pm 5.8 \right)~\mbox{MeV}\;,
\end{equation}

The value in (\ref{fbexp2}) is in good agreement with the data of
theoretical SM prediction \cite{Gray:2005ad}:
\begin{equation*}\label{fbsm}
f_{B}^P=\left(216.0 \pm 22.0 \right)~\mbox{MeV}\;,
\end{equation*}
but it is far enough from the data of \texttt{BaBar} and
\texttt{Belle} collaborations
\cite{Adachi:2008ch,Aubert:2009rk,Aubert:2007xj}, whose values lie
within the range $f_{B,\mathrm{exp}}^P= \left(246.8 \pm
45.6\right) ~\mbox{MeV}$. Value (\ref{fbexp1}) is closer to the
\texttt{BaBar} and \texttt{Belle} data.

Thus, on the base of existing uncertainties and experimental data
on the leptonic constants of heavy mesons, one can suppose that
$\alpha_{s}$- constant behavior has to freezing modes with
$\texttt{N}_{\alpha}=\texttt{5-{a}}-\texttt{7-{a}}$ and
$\alpha_{\mathrm{crit.}}= (0.660 \div 0.844 )$ as well as those
with \texttt{6-b}, \texttt{7-b} (can not definitely exclude the
behavior $\texttt{N}_{\alpha}=\texttt{4-\rm{a}}$, where
$\alpha_{\mathrm{crit.}}= 1.087$ and modes
\texttt{3-{b}}--\texttt{5-{b}}).

Note that bound state approach does not allow for the unique
identification of  $\alpha_{s}$ behavior, since the parameter
${A}\left(\mu\right)$, which this method is sensitive to, is
related with the latter indirectly through integration over the
momentum transfer $Q$.

Indeed, as follows from Tables \ref{tabfit} and \ref{tabfit1}, the
values of ${A}\left(\mu\right)$ for qualitatively different
regimes with freezing and peak are close to each other and lie
within the range of 0.13-0.19. These values of
${A}\left(\mu\right)=0.13-0.19$  are quite consistent with the
experimental data from \cite{Dokshitzer:1999sh}:
${A}_{\mathrm{fit}}\left(\mu\right)=0.16 \pm 0.01(\mathrm{exp})\pm
0.02(\mathrm{th})$.

To clarify the behavior of the effective coupling constant one
needs to analyze the experimental results from \texttt{SLAC}
\cite{Abe:1997cx,Abe:1998wq} and \texttt{JLab}
\cite{Amarian:2002ar,Fatemi:2003yh,Deur:2004ti,Dharmawardane:2006zd,Prok:2008ev}
on the first moments  of spin structure functions
$g^{p,n}_{1}(x,Q^2)$.

\section{First moments of spin structure functions \label{Sect7a}}

The first moments of spin-dependent proton and neutron structure
functions $g_1^{p,n}\left(x,Q^2\right)$ are defined as
\begin{equation}
\Gamma^{p,n}_{1}(Q^2)=\int\limits_{0}^{1}g_1^{p,n}\left(x,Q^2\right)\mathrm{d}x\;
.\label{gpn}
\end{equation}

The pQCD result  (\ref{gpn}) (in the
$\overline{\mathrm{MS}}$-scheme) is
\begin{equation}
\Gamma^{p,n}_{1}(Q^2)=\left\{\frac{1}{12}\left(\pm g_A
+\frac{a_8}{3}\right) C_{NS}\left(Q^2\right)+\frac{{a}_0^{inv}}{9}
C_{SI}^{inv}\left(Q^2\right)\right\}+
\Delta_{\mathrm{HT}}^{p,n}(Q^2)\; ,\label{gpn1}
\end{equation}
where $C_{NS}\left(Q^2\right)$ and $C_{SI}^{inv}\left(Q^2\right)$
are the first moments of the non-singlet and singlet Wilson
coefficient functions, respectively. Here the $+(-)$ sign of the
$g_A$ term holds for the proton (neutron).

The coefficients $C_{NS}$ and $C_{S}^{inv}$ have been calculated
in perturbative QCD (in the $\overline{\mathrm{MS}}$ scheme) up to
the third and fourth order in $\alpha_{s}$ (see
\cite{Larin:1994dr,Kataev:1994gd,Larin:1997qq,Baikov:2010je})
\begin{equation}
C_{NS}\left(Q^2\right)= 1+\sum\limits_{k=1}^{4} d^{NS}_{k} a_s^k(Q^2)\; ,\label{cns}
\end{equation}
\begin{equation}
C_{SI}^{inv}\left(Q^2\right)= 1+\sum\limits_{k=1}^{3} d^{SI}_{k}a_s^k(Q^2)\; ,\label{csi}
\end{equation}
where
\begin{eqnarray}
a_s(Q^2) &=& \alpha_s(Q^2)/\pi \;,\nonumber \\
d_1^{NS} &=& -1\;,\nonumber \\
d_2^{NS} &=& -4.583 + 0.333\,n_f\; ,\nonumber \\
d_3^{NS} &=&-41.44 + 7.607\,
n_f-0.1775\, n_f^2 \; , \nonumber \\
d_4^{NS} &=& -479.4475+123.3914\, n_f-7.6975\, n_f^2+0.10374\,
n_f^3\; ,\label{dns}
\end{eqnarray}
and
\begin{eqnarray}
d_1^{SI} &=& \frac{1}{4 \beta_{0}}\left(-11+\frac{8}{3} \,n_f\right)\;,\nonumber \\
d_2^{SI} &=& \frac{1}{4^{2}\beta_{0}^2}\left(-554.583 +214.466 \,n_f-16.4757\, n_f^2+0.5537 \,n_f^3\right)\; ,\nonumber \\
d_3^{SI} &=& \frac{1}{4^{3}\beta_{0}^3}\left(-55156.5+29692.9\,
n_f
- 5292.89 \,n_f^2+434.2 \,n_f^3-\right.\nonumber \\
&-& \left. 16.139 \,n_f^4+0.229263 \,n_f^5\right) \;. \label{ds}
\end{eqnarray}

Constants $g_A = 1.2701 \pm 0.0025$ \cite{pdg2012} and $a_8 =
0.585 \pm 0.025$ \cite{Leader:2002az} are the isovector and
$SU(3)$ octet axial charges, respectively. The ${a}_0^{inv}$ is
the renormalization group invariant (i.e. $Q^2$ independent)
\cite{Larin:1994dr,Larin:1997qq}.

The  $Q^2$ evolution of the axial singlet charge $a_{0}(Q^2)$
\cite{Kataev:1994gd,Larin:1997qq} for $n_f=3$ quarks flavors is
\begin{eqnarray}
a_{0}(Q^2)&=& {a}_0^{inv} \exp\left(\int\limits^{a_s}\frac{\gamma
SI(x)}{\beta(x)}\mathrm{d}x \right) \approx
\nonumber \\
&\approx&\left( 1 + \frac{2}{3} a_s(Q^2)+ \frac{131}{108} a_s^2(Q^2) + \frac{41477}{11664}
a_s^3(Q^2) \right) {a}_0^{inv}\; .\label{a0q2}
\end{eqnarray}

The product ${a}_0^{inv} C_{SI}^{inv}\left(Q^2\right)$ is often
rewritten in the form
\begin{equation}
{a}_0^{inv}
C_{SI}^{inv}\left(Q^2\right)={a}_0(Q^2){C}_{SI}\left(Q^2\right)\;
,\label{csi2}
\end{equation}
where ${C}_{SI}\left(Q^2\right)$  up to the second order
$\alpha_{s}$ and $n_f=3$ is determined by
\begin{equation}
{C}_{SI}\left(Q^2\right)=1 - a_s(Q^2)- 1.096\;
a_s^2(Q^2)+\mathcal{O}\left(\alpha_{s}^{3}\right)\; .\label{csi2a}
\end{equation}

Interesting to note that analytical expressions for $C_{NS}$ and
${C}_{SI}$ , defined in Eq. (\ref{gpn1}), are identical in all
orders of perturbation theory in the conformal invariant limit of
the massless $SU(N_c)$ gauge model with fermions
\cite{Kataev:2012kc}.

The second term $\Delta_{\mathrm{HT}}(Q^2)$ in the Eq.(\ref{gpn1})
is a contribution of higher twists
\begin{equation}
\Delta_{\mathrm{HT}}^{p,n}(Q^2)=\sum\limits_{i=2}^{\infty}\frac{\mu_{2i}^{p,n}\left(Q^2\right)}{Q^{2i-2}}\;
.\label{twist}
\end{equation}

If the expression (\ref{gpn1}) uses ``frozen'' constants, it is
necessary to modify the argument of the $\mathrm{HT}$-function by
replacement \cite{Badelek1997zx,2005JETPKotikov,Shirkov2013mrg}
\begin{equation}
Q^2 \to Q^2 + m_{ht}^{2} \; .\label{q2tw}
\end{equation}
Therefore, we have the following form for the function
(\ref{twist})
\begin{equation}
\Delta_{\mathrm{HT}}^{p,n}(Q^2) \to
\Delta_{\mathrm{HT}}^{p,n}(Q^2,m_{ht})=
\sum\limits_{i=2}^{\infty}\frac{\mu_{2i}^{p,n}\left(Q^2\right)}{\left(Q^2+m_{ht}^2\right)^{i-1}}\;
.\label{twist2}
\end{equation}
As noted in \cite{Shirkov2013mrg} the value $m_{ht}$  is close to
the $m_g$ one.

There are relations, describing the dependence of $Q^2$ for the
$\mu_4$
\begin{equation}
\mu_4\left(Q^2\right)=\mu_4\left(Q_{0}^2\right)
\left(\frac{\alpha_s\left(Q^2\right)}{\alpha_s\left(Q_{0}^2\right)}\right)^{8/(9
\beta_0)} \; ,\label{mu4q2}
\end{equation}
while the $Q^2$-evolution of the higher twists $\mu_{6}, \mu_{8}$
is theoretically unknown (see,
\cite{Deur:2005jt,Pasechnik:2009yc,Pasechnik:2010fg}).

Difference functions (\ref{gpn1}) $\Gamma^{p,n}_{1}$ lead to the
QCD-modified Bjorken sum rule (BSR)
\cite{Bjorken:1966jh,Bjorken:1969mm}
\begin{equation}
\Gamma^{p-n}_{1}(Q^2)\equiv\Gamma^{p}_{1}(Q^2)-\Gamma^{n}_{1}(Q^2)=
\frac{g_A}{6}
C_{NS}\left(Q^2\right)+\Delta_{\mathrm{HT}}^{p-n}\left(Q^2\right)
\;.\label{bsum}
\end{equation}

\section{Determination of the ``optimal''  $\texttt{N}_{\alpha}$ from BSR \label{bsr}}

At present we have an extensive experimental information about the
first moments $\Gamma^{p,n}_{1}(Q^2)$
\cite{Abe:1997cx,Abe:1998wq,Adeva:1998vw,Airapetian:2007mh,Airapetian:2002wd,Amarian:2002ar,Fatemi:2003yh,Deur:2004ti,Dharmawardane:2006zd,Prok:2008ev}.
This data allow us to study the behavior of the effective coupling
constants in the nonperturbative region and to clarify their
possible behavior through Bjorken (\ref{bsum}) and Ellis-Jaffe
(\ref{gpn1}) sum rules.  In Refs.
\cite{Deur2009,Khandramai:2013haz,Kotikov2013,Pasechnik:2009yc,Pasechnik:2010fg,Shirkov2013mrg}
this information is used to analyze the behavior of effective QCD
constants at low $Q^2$.

First, let is consider  function (\ref{bsum})  as the best
agreement with the experimental data $\Gamma^{p-n}_{1,\,
\mathrm{exp}}(Q^2)$ in terms of the behavior $\alpha_{s}$
constant. We assume that the contribution of higher twists to
(\ref{bsum}) equal to zero and the number of active flavours
$n_f=3$. The optimal behavior of effective strong constant is
obtained by finding the minimum function
\begin{equation}
\label{chi2gpn} \chi^{2}\left(\texttt{N}_{\alpha}\right)= \sum_{i=
1}^{k} \frac{\left( \Gamma^{p-n}_{1,\,
\mathrm{exp}}(Q^2_{i})-\Gamma^{p-n}_{1}(Q^2_{i}\right)^{2}}{
\left(\delta
\Gamma^{p-n}_{1,\,\mathrm{exp}}(Q^2_{i})\right)^{2}}\; .
\end{equation}
Here the errors $\delta \Gamma^{p-n}_{1,\,\mathrm{exp}}(Q^2_{i})$
are ones for   \texttt{JLab}
\cite{Amarian:2002ar,Fatemi:2003yh,Deur:2004ti,Dharmawardane:2006zd,Prok:2008ev}
and \texttt{SLAC} \cite{Abe:1997cx,Abe:1998wq} data sets except
for the data of \texttt{HERMES}
\cite{Airapetian:2007mh,Airapetian:2002wd}.

\begin{figure}[h t b p]
\begin{center}
\vspace{0mm} \resizebox{0.7\textwidth}{!} {
\includegraphics{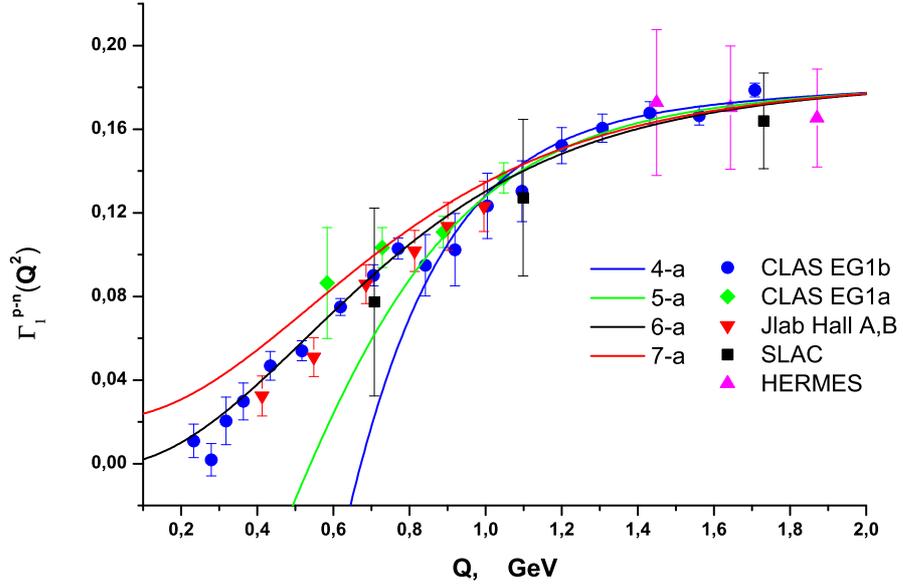}}
\vspace{-5mm}
\end{center}
\caption{ \texttt{Jlab}, \texttt{SLAC} and \texttt{HERMES}
experimental data for BSR. The curves represent model predictions
obtained for modes with freezing constant
$\texttt{N}_{\alpha}=\texttt{4-a}-\texttt{7-a}$ without higher
twists (\ref{twist2}).} \label{fig7}
\end{figure}

\begin{figure}[h t b p]
\begin{center}
\vspace{-5mm}
\resizebox{0.7\textwidth}{!} {
\includegraphics{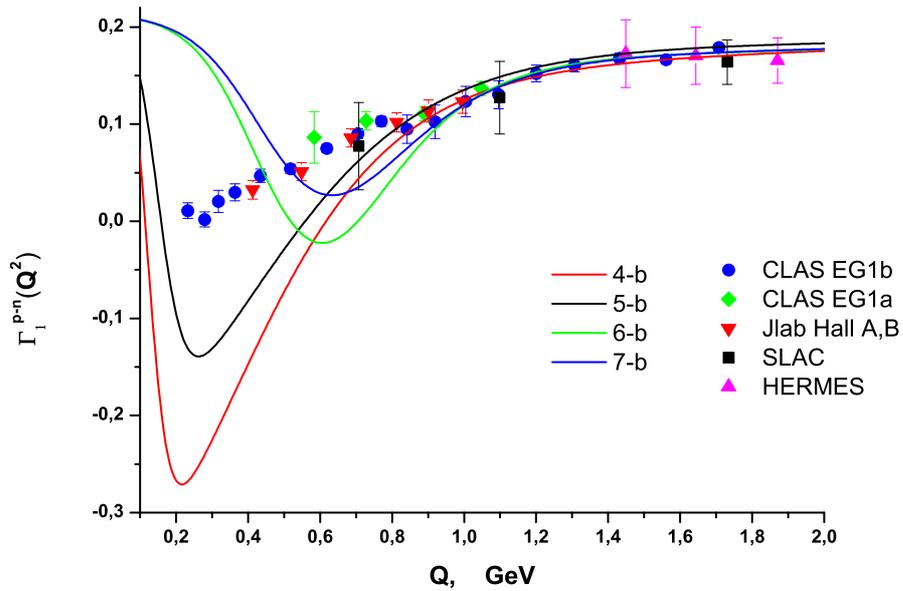}}
\vspace{-5mm}
\end{center}
\caption{Same as Fig.\ref{fig7}, but for constants
$\texttt{N}_{\alpha}=\texttt{4-b}-\texttt{7-b}$ with a maximum in
the nonperturbative region.} \label{fig8}
\end{figure}

The results of calculations with experimental errors are presented
in Table \ref{chi2alfagpn} and Figs. \ref{fig7},\ref{fig8}.
\begin{table}[h t ]
\begin{center}
\caption{ The values of
$\chi^{2}\left(\texttt{N}_{\alpha}\right)$/D.o.f. for various
regimes of $\alpha_{s}$ behavior. \label{chi2alfagpn}}
\vspace{2mm}
\begin{tabular}{|c|c||c|c|}
  \hline
  $\texttt{N}_{\alpha}$ & $\chi^{2}\left(\texttt{N}_{\alpha} \right)$/D.o.f. &
  $\texttt{N}_{\alpha}$ & $\chi^{2}\left(\texttt{N}_{\alpha}\right)$/D.o.f\\
  \hline
  \texttt{4-a} & $515.1$  &\texttt{4-b}  & $170.9$  \\
  \texttt{5-a} & $47.4 $  &\texttt{5-b}  & $61.5$   \\
  \texttt{6-a} & $0.7  $  &\texttt{6-b}  & $84.8$   \\
  \texttt{7-a} & $3.3  $  &\texttt{7-b}  & $59.7$   \\
  \hline
\end{tabular}
\end{center}
\end{table}

The first conclusion that follows from the calculations is the
following: the  behavior, in which constants $\alpha \to 0$ when
$Q^2 \to 0$ does not properly describe the experimental data. The
highest acceptance probability  (minimum $\chi^{2}$) are constants
for sets $\texttt{N}_{\alpha}=\texttt{5-{b}}$ and
$\texttt{N}_{\alpha}=\texttt{7-{b}}$. Accounting for higher twists
does not change this conclusion as well.

As follows from the calculations (see Table \ref{chi2alfagpn}),
models with freezing constant $\texttt{N}_{\alpha}\hm
=\texttt{6-{a}}, \texttt{7-{a}}$ have a minimum of $\chi^{2}$.
From Figure \ref{fig7} one can see that mode
$\texttt{N}_{\alpha}\hm =\texttt{6-{a}}$  allows us to describe
the Bjorken sum rule data.

\begin{figure}[h t b p]
\begin{center}
\vspace{-5mm} \resizebox{0.7\textwidth}{!} {
\includegraphics{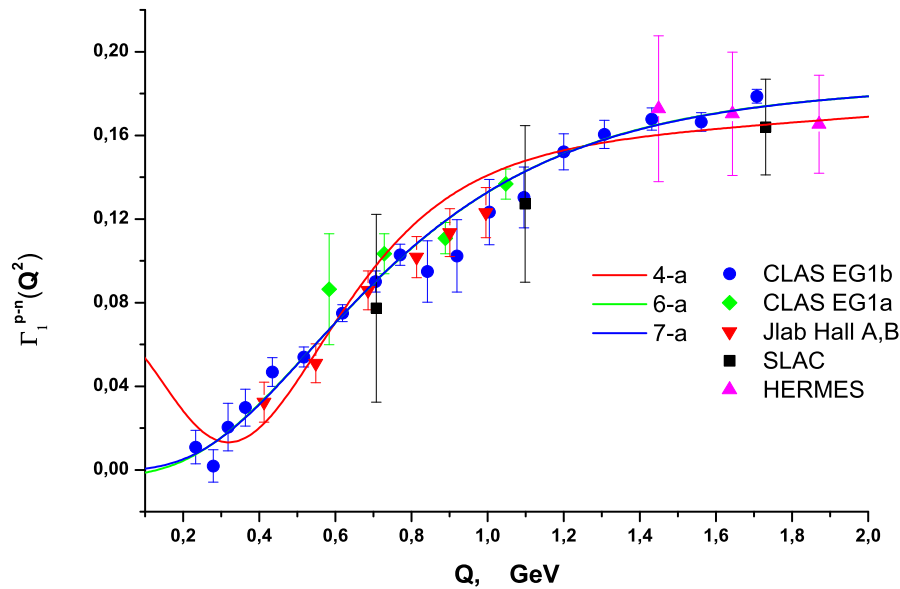}}
\vspace{-5mm}
\end{center}
\caption{ The curves represent model predictions obtained for
modes with freezing constant $\texttt{N}_{\alpha}=\texttt{4-a},
\texttt{6-a}-\texttt{7-a}$, including higher twists (\ref{twist2})
and $m_{ht}=0.6 \mbox{~GeV}$}. \label{fig9}
\end{figure}

\begin{figure}[h t b p]
\begin{center}
\vspace{-5mm} \resizebox{0.7\textwidth}{!} {
\includegraphics{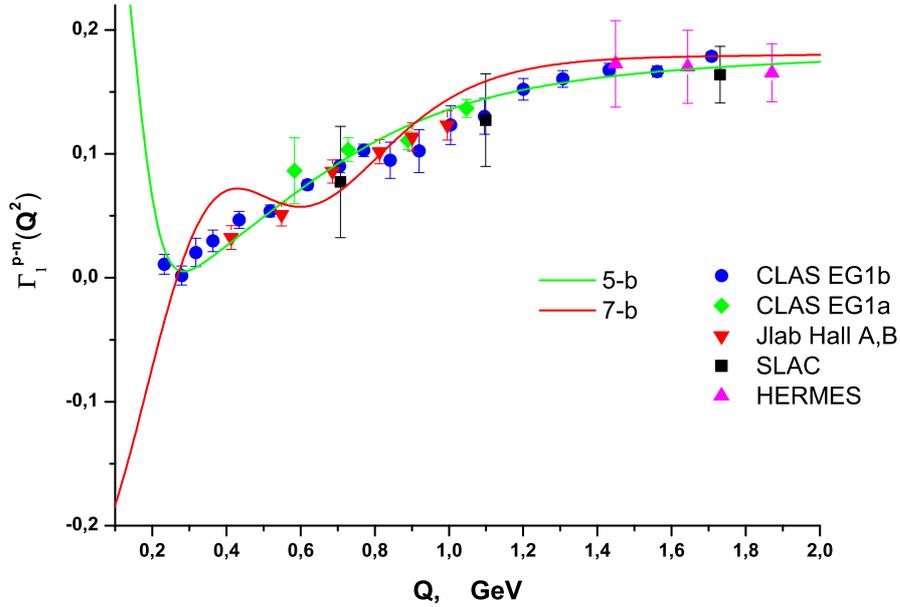}}
\vspace{-5mm}
\end{center}
\caption{Same as Fig.\ref{fig9}, but for constants
$\texttt{N}_{\alpha}=\texttt{5-b}$ and
$\texttt{N}_{\alpha}=\texttt{7-b}$ with a maximum in the
nonperturbative region.} \label{fig10}
\end{figure}

Let's briefly consider the contributions of higher-twist  to the
above conclusions. The results of calculations are given in Table
\ref{chi2alfagp} and Figs. \ref{fig9}, \ref{fig10}.
\begin{table}[h t ]
\begin{center}
\caption{Dependence of the
$\chi^{2}\left(\texttt{N}_{\alpha}\right)$/D.o.f. on the
$3$-parametrers fit results of BSR data with $m_{ht}=0.6
\mbox{~GeV}$. The corresponding fit curves are shown in Figs.
\ref{fig9} and \ref{fig10}. \label{chi2alfagp}} \vspace{2mm}
\begin{tabular}{|c|c||c|c|}
  \hline
  $\texttt{N}_{\alpha}$ & $\chi^{2}\left(\texttt{N}_{\alpha} \right)$/D.o.f. &
  $\texttt{N}_{\alpha}$ & $\chi^{2}\left(\texttt{N}_{\alpha}\right)$/D.o.f\\
  \hline
  \texttt{4-a} & $2.1 $  &\texttt{4-b}   & $1.3$  \\
  \texttt{5-a} & $0.7 $  &\texttt{5-b}   & $0.9$   \\
  \texttt{6-a} & $0.5 $  &\texttt{6-b}   & $8.4$   \\
  \texttt{7-a} & $0.5$   &\texttt{7-b}   & $5.4$   \\
  \hline
\end{tabular}
\end{center}
\end{table}
As follows from the calculations, higher twists  can improve the
agreement between the experimental data and model calculations.
However, the smallest $\chi^{2}$ are again constants for
$\texttt{N}_{\alpha}=\texttt{6-a}-\texttt{7-a}$.

With the help of (\ref{bsum}), (\ref{cns}) and (\ref{dns}) it is
easy to estimate the critical value of the strong coupling
constant $\alpha_{\mathrm{crit.}}$.  If we assume that the value
$\Gamma_{1,\mathrm{exp}}^{p-n}(Q^2) \to 0$ when $Q^2 \to 0$, and
the contribution of higher twists
$\Delta_{\mathrm{HT}}^{p-n}\left(Q^2=0\right)$ is less than $\sim
0.1$, then, after solving equation
\begin{equation}
C_{NS}\left(Q^2=0\right)= 0 \pm
\Delta_{\mathrm{HT}}^{p-n}\left(Q^2=0\right)\label{cnscrit}
\end{equation}
we find the following numerical estimates
\begin{equation}
\alpha_{\mathrm{crit.}}= 0.6861 \pm 0.0257\; . \label{critgpn}
\end{equation}
This value is in excellent agreement with constants with
$\texttt{N}_{\alpha}=\texttt{6-a}$ and
$\texttt{N}_{\alpha}=\texttt{7-a}$ modes.

In this paper, we do not plan to perform  detailed calculation of
the contributions of higher twists. Note that the fitting
procedure shows a strong correlation between the coefficients
$\mu_{4},\mu_{6}$ and $\mu_{8}$. Therefore, it is difficultly to
reach a clear estimation  of the contributions of each of the
terms in (\ref{twist2}). The solution of this problem is to either
reduce the number of parameters to one in
$\Delta_{\mathrm{HT}}^{p-n}\left(Q^2\right)$ (only $\mu_{4}$) or
to search for additional conditions, which limit the values of
coefficients $\mu_{4},\mu_{6}$ and $\mu_{8}$.

\section{Determination of the ``optimal''  $\texttt{N}_{\alpha}$ from $\Gamma_{1}^{p,n}$ \label{Sect8a}}

In this section, we perform a similar procedure for computing
estimates of section \ref{bsr} using the experimental data for the
first moment $\Gamma_{1}^{p,n}$.

In order to find the optimal behavior of the simulated mode, we
use the combined
$\chi^{2}_{\mathrm{com}}\left(\texttt{N}_{\alpha}\right)$:
\begin{equation}
\chi^{2}_{\mathrm{com}}\left(\texttt{N}_{\alpha}\right)=
\chi^{2}_{p}\left(\texttt{N}_{\alpha}\right)/D.o.f+\chi^{2}_{n}\left(\texttt{N}_{\alpha}\right)/D.o.f\;
,\label{chitot}
\end{equation}
where
\begin{equation}
\label{chi2gpn1} \chi^{2}_{p,n}\left(\texttt{N}_{\alpha}\right)=
\sum_{i= 1}^{k} \frac{\left( \Gamma^{p,n}_{1,\,
\mathrm{exp}}(Q^2_{i})-\Gamma^{p,n}_{1}(Q^2_{i}\right)^{2}}{
\left(\delta
\Gamma^{p,n}_{1,\,\mathrm{exp}}(Q^2_{i})\right)^{2}}\; .
\end{equation}

\begin{figure}[h t b p]
\begin{center}
\vspace{0mm} \resizebox{0.7\textwidth}{!} {
\includegraphics{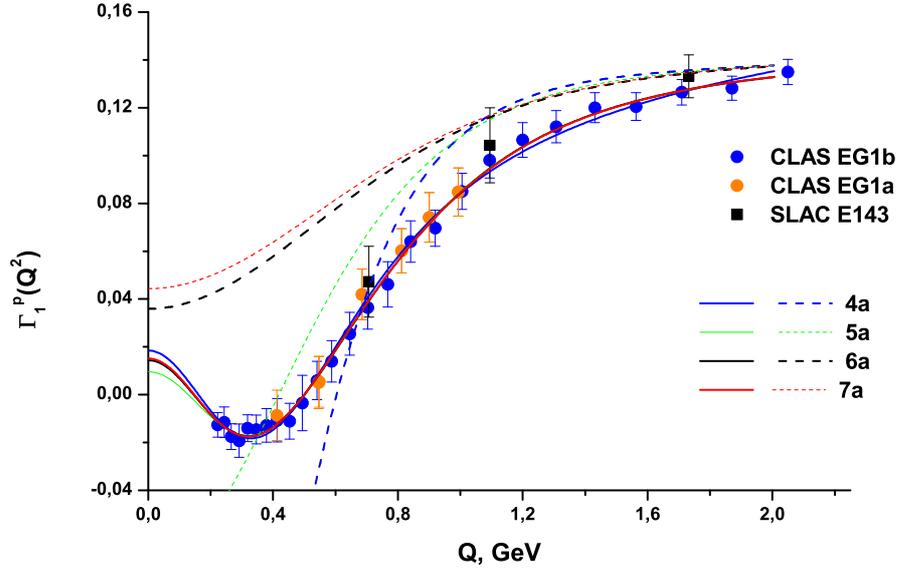}}
\vspace{0mm}
\end{center}
\caption{The curves represent model predictions obtained for modes
with freezing constant
$\texttt{N}_{\alpha}=\texttt{4-a}-\texttt{7-a}$ with (solid lines)
and without (dashed lines) higher twists (\ref{twist2}) for
$m_{ht}=0.6 \mbox{~GeV}$.} \label{fig11}
\end{figure}

\begin{figure}[h t b p]
\begin{center}
\vspace{0mm} \resizebox{0.7\textwidth}{!} {
\includegraphics{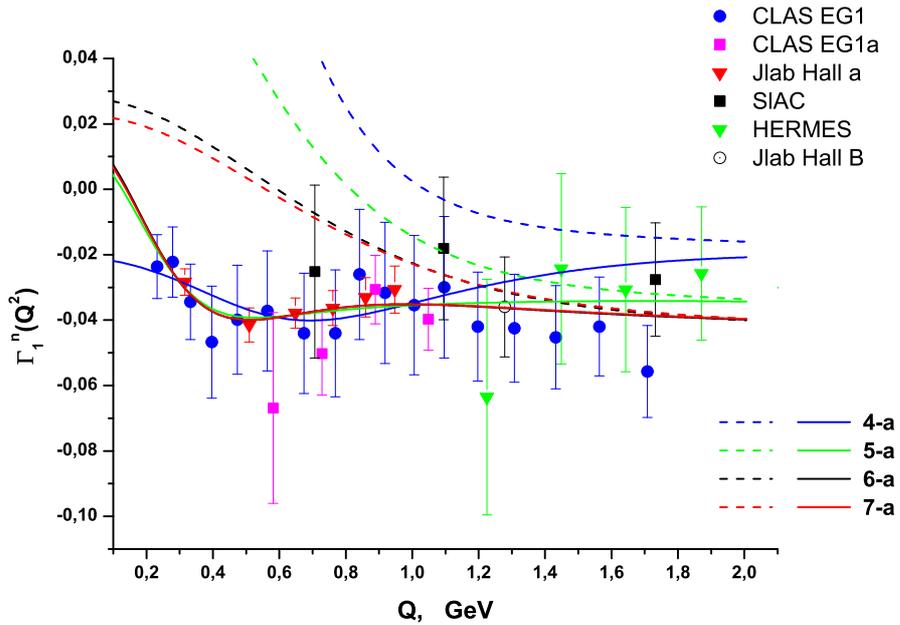}}
\vspace{-5mm}
\end{center}
\caption{Same as Fig.\ref{fig11}, but for
$\Gamma_{1}^{n}(Q^{2})$.} \label{fig12}
\end{figure}

As follows from the calculations (see Figs. \ref{fig11},
\ref{fig12}), to describe the behavior of
$\Gamma^{p,n}_{1,\,\mathrm{exp}}(Q^{2})$ one must take into
account higher twists, while the contribution $\Delta^{p-n}$ of
BSR can be almost neglected (mode
$\texttt{N}_{\alpha}=\texttt{6-a}$).
The corresponding results for
$\chi^{2}_{\mathrm{com}}\left(\texttt{N}_{\alpha}\right)$ with
higher twists and without them are listed in Table
\ref{chi2alfagp2}.
\begin{table}[h t p b]
\begin{center}
\caption{Dependence of the
$\chi^{2}_{com}\left(\texttt{N}_{\alpha}\right)$/D.o.f. on the
$4$-parameters fit results of $\Gamma_{1}^{p,n}(Q^2)$ data with
$m_{ht}=0.6 \mbox{~GeV}$. The corresponding fit curves are shown
in Figs. \ref{fig11} and \ref{fig12}. \label{chi2alfagp2}}
\vspace{2mm}
\begin{tabular}{|c|c|c|}
  \hline
  $\texttt{N}_{\alpha}$ & $\chi^{2}_{com}\left(\texttt{N}_{\alpha} \right)$/D.o.f.&
   $\chi^{2}_{com}\left(\texttt{N}_{\alpha}\right)$/D.o.f \\
  &  without HT terms    & with HT terms  \\
  \hline
  \texttt{4-a} & $482.8$     &  $0.78 $ \\
  \texttt{5-a} & $65.1 $     &  $0.43 $  \\
  \texttt{6-a} & $52.9 $     &  $0.39 $  \\
  \texttt{7-a} & $63.8 $     &  $0.40$   \\
  \hline
\end{tabular}
\end{center}
\end{table}

The parameter $m_{ht}=0.6 \mbox{~GeV}$ is chosen so that the
result of fitting to the proton $\Gamma_{1}^{p}(Q^2)$ data for
axial charge
\begin{equation}
a_0^{inv}=0.325 \pm 0.063 \; ,\label{aoinvfit}
\end{equation}
is in good agreement with the analysis of \texttt{COMPASS} group
\cite{Alexakhin:2006vx}
$$a_0^{inv}=0.33 \pm 0.03 (\mbox{stat.}) \pm 0.05 (\mbox{syst.}).$$

The  $Q^2$ evolution of the axial singlet charge $a_{0}(Q^2)$ can
now be obtained from  Eq.(\ref{a0q2}) using Eq.(\ref{aoinvfit})
\begin{eqnarray}
&& {a}_0(Q^2=3 \mbox{~GeV}^2)= 0.353 \pm 0.069\; , \nonumber \\
&& a_0(Q^2=5 \mbox{~GeV}^2)=  0.349 \pm 0.067\;\; . \label{a0q2a}
\end{eqnarray}
Also, the results (\ref{a0q2a}) and experimental values of
\texttt{COMPASS} \cite{Alexakhin:2006vx}
$${a}_0(Q^2=3 \mbox{~GeV}^2)= 0.35  \pm 0.03
(\mbox{stat.}) \pm 0.05 (\mbox{syst.})$$ and \texttt{HERMES}
\cite{Airapetian:2007mh} groups
$${a}_0(Q^2=5 \mbox{~GeV}^2)= 0.330 \hm \pm 0.011(\mathrm{theo.}) \pm
0.025(\mathrm{exp.}) \pm 0.028(\mathrm{evol.})$$ are consistent
with each other.

\begin{figure}[h t b p]
\begin{center}
\vspace{0mm} \resizebox{0.7\textwidth}{!} {
\includegraphics{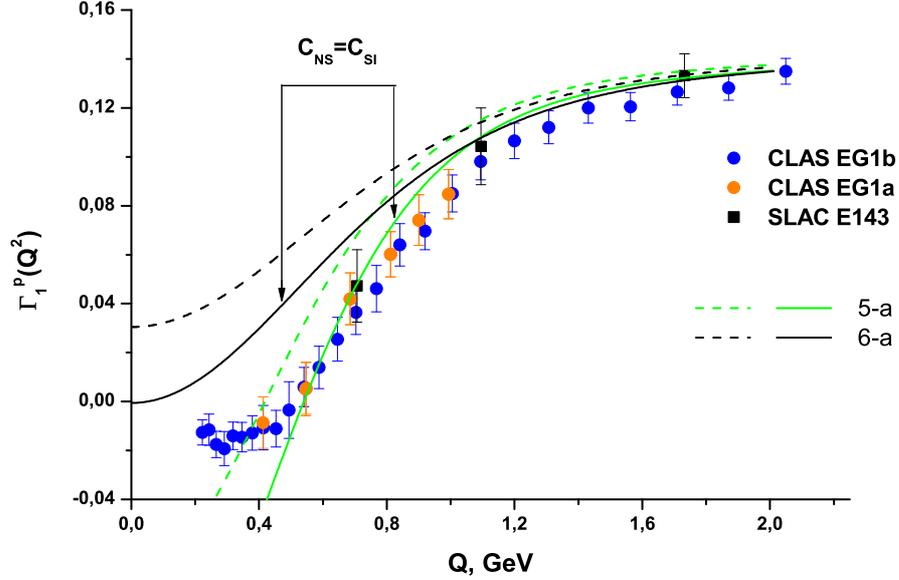}}
\vspace{0mm}
\end{center}
\caption{ The curves represent the model predictions (without
higher twists) $\Gamma_{1}^{p}$ obtained for modes with freezing
constant $\texttt{N}_{\alpha}=\texttt{5-a}-\texttt{6-a}$,  when
$C_{NS}=C_{SI}$ (solid lines) and  (dashed lines) when $C_{NS}$, $
C_{SI}^{inv}$ are determined by (\ref{cns}), (\ref{dns}) and
(\ref{csi}), (\ref{ds}).} \label{fig13}
\end{figure}

It is interesting to note that the assumption of conformal
invariance \cite{Kataev:2012kc} improves the agreement between
model predictions and experimental data (see Fig. \ref{fig13}). It
can serve as an indirect argument for existence of conformal
invariant limit of the massless $SU(N_c)$ gauge model.

\section{``Optimal''  $\texttt{N}_{\alpha}$ from GLS \label{Sect9a}}

The Gross-Llewellyn Smith (GLS) sum rule \cite{Gross:1969jf} predicts the integral
\begin{equation}
{K}_{GLS}\left(Q^2\right)=
\frac{1}{2}\int\limits_{0}^{1}F_3\left(x,Q^2\right)\mathrm{d}x =
3\; {C}_{GLS}\left(Q^2\right)\;, \label{GLS}
\end{equation}
where $x F_3(x,Q^2)$ is the nonsinglet structure function measured
in $\nu N$-scattering (see \cite{Hinchliffe:1996hc,Kim:1998kia}
and references therein).

The calculation  of the $\mathcal{O}\sim \alpha_{s}^{4}$ contribution to $C_{GLS}$  has been
published in \cite{Baikov:2010iw,Baikov:2012zn} and the function $C_{GLS}$ can be written in the
form
\begin{equation}
{C}_{GLS}\left(Q^2\right)= C_{NS}\left(Q^2\right)+  C_{SI}^{GLS}\left(Q^2\right)\; ,\label{gls}
\end{equation}
where the function $C_{NS}\left(Q^2\right)$ is determined by Eq.(\ref{csi}) and (\ref{ds}). Singlet
contribution is defined by
\begin{equation}
C_{SI}^{GLS}\left(Q^2\right)=0.4132 \; n_f \; a_s^3(Q^2) +n_f\;
(5.802-0.2332 \; n_f) a_s^4(Q^2) \; .\label{gls-si}
\end{equation}

\begin{figure}[h t b p]
\begin{center}
\vspace{0mm} \resizebox{0.7\textwidth}{!} {
\includegraphics{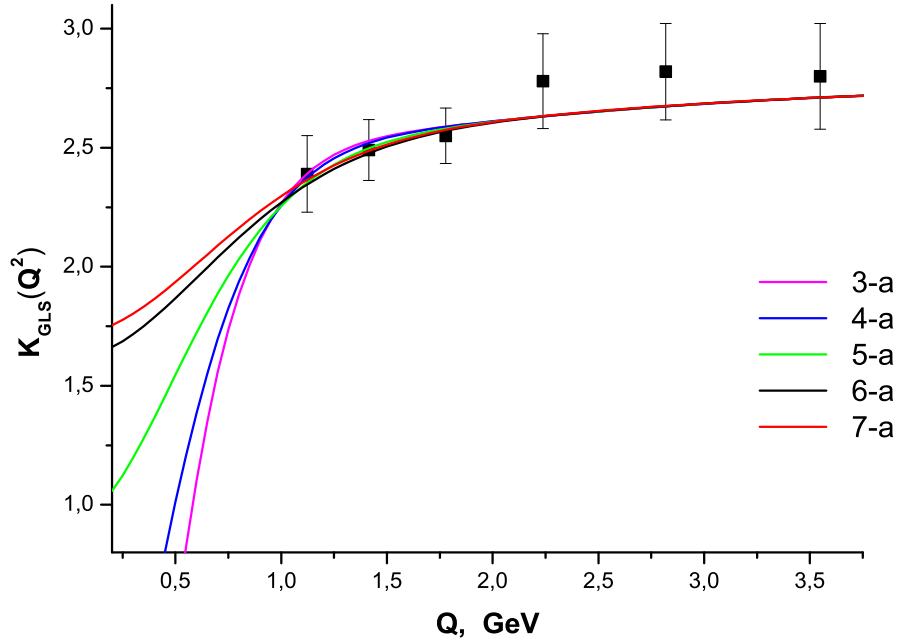}}
\vspace{0mm}
\end{center}
\caption{pQCD results of $K_{GLS}$ for different freezing variants
of $\alpha_{GI}$.} \label{fig14}
\end{figure}

Let us compare pQCD results of $K_{GLS}$  with relevant
experimental data  \cite{Kim:1998kia}. We plot the data in Fig.
\ref{fig14} along with model predictions for different variants of
$\alpha_{GI}$.

Within the considerable error bars we see that different versions
of $\alpha_{s}$ (\ref{alsAn}) and expe\-rimental data are well
compatible with each other in order. To distinguish the behaviour
of constants  experimental data at lower $Q^2 < 1 \mbox{~GeV}$ is
needed.

Thus, modern experimental data on the GLS sum rule does not allow
to determine  the mode of QCD constants behavior.

\section{Conclusion \label{Sect11a}}

In this paper a method  that allows the behavior of the QCD
running coupling constant to be assessed in the nonperturbative
region is proposed. To study the probable behavior of the QCD
constant, 14 regimes were simulated with different
$\alpha_{\mathrm{crit.}}=\alpha_{s}\left(0\right)$ and $Q^2$
behavior in the nonperturbative region (Tables \ref{tabfit} and
\ref{tabfit1}).

The requirements that underline this method are the matching
condition between calculations, which are done within the
framework of relativistic quark model (the Poincar\'{e}-covariant
model) with interquark potential (\ref{pvpotential}), and
experimental data on the masses and constants of leptonic decays
of pseudoscalar and vector mesons.

Based on the analysis, the compliance with model calculations of
the experimental information on the leptonic constants of heavy
($B, D,$ and $D_s$) mesons and sum rules of the nucleon can be
confirmed. Most appropriate mode is freezing constant with the
critical value $\alpha_{\mathrm{crit.}}$ is between $0.65-0.72$
(regimes $\texttt{N}_{\alpha}=\texttt{6-a}$ and
$\texttt{N}_{\alpha}=\texttt{7-a}$).

Let us consider what model of freezing strong constant has a
similar behavior at small $Q$. These models are:
\begin{itemize}
    \item MPT coupling constant (\ref{mpt}) with $m_g=1.0  \mbox{~GeV}$,
    $\alpha_{\mathrm{crit.}}=0.72$ and
    $\Lambda^{\left(n_f=3\right)}\hm =
    380 \mbox{~MeV}$ \cite{Shirkov:1999hm,Shirkov2013mrg}.
    \item BPT coupling constant (\ref{alhasbpt}) with $m_g=1.08  \mbox{~GeV}$,
    $\alpha_{\mathrm{crit.}}=0.72$ and $\Lambda^{\left(n_f=3\right)}\hm =507
    \mbox{~MeV}$ \cite{Badalian:2001by}.
    \item The nonperturbative effective coupling
    (\ref{conwallconst}) with $m_g=0.34\mbox{~GeV}$,
    $\alpha_{\mathrm{crit.}}\hm = 0.71$ and $\Lambda^{\left(n_f=3\right)}\hm =256
    \mbox{~MeV}$ \cite{Cornwall:1981zr}.
   \item The coupling constant $\alpha_{\rm W}^{(1)} ( Q^{2} )$ (\ref{weberconst}) \cite{Webber:1998um} with parameters
   $c=p=1.71$, $d=0.54$ and $\Lambda^{\left(n_f=3\right)}=250
    \mbox{~MeV}$.
\end{itemize}

\begin{figure}[h t]
\begin{tabular}{cc}
\subfigure[Model predictions of $\Gamma^{p-n}_{1}(Q^2)$ obtained
for modes with freezing constants: $\alpha_{\rm MPT}$ (\ref{mpt}),
$\alpha_{\rm BPT}$ (\ref{alhasbpt}), $\alpha_{\rm Con}$
(\ref{conwallconst}), $\alpha_{\rm Web}$ (\ref{weberconst}) and
mode $\texttt{N}_{\alpha}=\texttt{6-a}$]{\includegraphics[scale
=0.75]{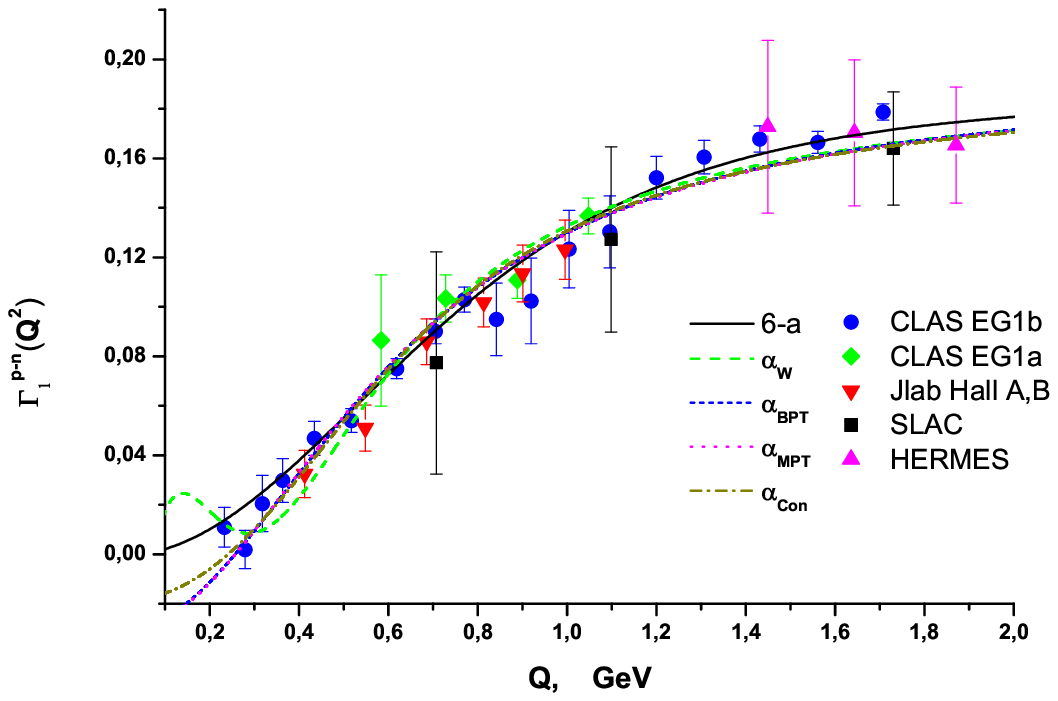}} &~
\subfigure[Effective running strong coupling constants:
$\alpha_{\rm MPT}$ (\ref{mpt}), $\alpha_{\rm BPT}$
(\ref{alhasbpt}), $\alpha_{\rm Con}$ (\ref{conwallconst}),
$\alpha_{\rm Web}$ (\ref{weberconst}) and mode
$\texttt{N}_{\alpha}=\texttt{6-a}$]{\includegraphics[scale
=0.75]{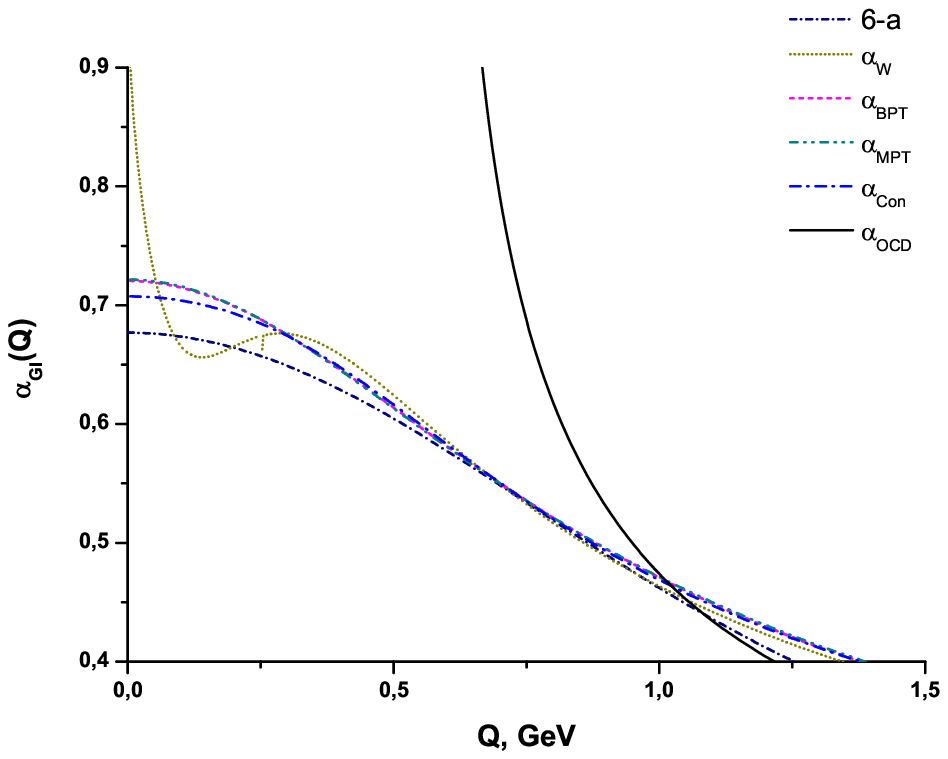}}
\end{tabular}
\caption{Model calculations of the Bjorken sum rule without higher
twist and nonperturbative behavior of the various models of
effective QCD constants.} \label{figalfas4}
\end{figure}

For comparison, the behavior of the effective constants in the
nonperturbative region for various approaches and the improved
para\-meteri\-zation of mode \texttt{6-{a}} are presented in Fig.
\ref{figalfas4}-(b) (right panel). Figure \ref{figalfas4}-(a)
shows model calculations of the Bjorken sum rule  without higher
twists. As we can see, all these models have is identical behavior
in the nonperturbative region, except for the constant
(\ref{weberconst}).

I  would like to thank O.P. Solovtsova, A.E. Dorokhov, and the
attendees of the seminar held at the Laboratory of Theoretical
Physics (JINR) for stimulating discussions and advisory comments.


\end{document}